\documentclass[a4paper,UKenglish,cleveref, autoref, thm-restate]{lipics-v2021}
\usepackage[utf8]{inputenc}
\usepackage{amsmath,graphicx,amsthm,dsfont}

\title{Multi-Dimensional Stable Roommates in $2$-Dimensional Euclidean Space}

\author{Jiehua Chen}{TU Vienna, Austria}{jiehua.chen@tuwien.ac.at}{}{Vienna Science and Technology Fund (WWTF) grant VRG18-012}
\author{Sanjukta Roy}{Faculty of Information Technology, Czech Technical University in Prague, Czech Republic;\\
TU Vienna, Austria}{sanjukta.roy@tuwien.ac.at}{}{This work was done when SR was affiliated with TU Vienna, and was supported by Vienna Science and Technology Fund (WWTF) grant VRG18-012}

\authorrunning{J. Chen and S. Roy} %

\Copyright{CC-BY} %



\category{} %

\nolinenumbers %
\usepackage{color,soul}
\usepackage{xspace}
\usepackage{enumerate}
\usepackage{mathtools} 

\usepackage{multicol}
\usepackage{booktabs}

\usepackage{subcaption}
\usepackage[noend,ruled,linesnumbered]{algorithm2e} 
\usepackage{xifthen}

\usepackage[textsize=tiny,textwidth=1.5cm,linecolor=green!70!black, backgroundcolor=green!10, bordercolor=black,disable=true]{todonotes}%

\usepackage{mdframed}

\usepackage{tikz}
\usetikzlibrary{decorations,arrows,petri,topaths,backgrounds,shapes,positioning,fit,calc,decorations.pathreplacing,patterns,intersections,decorations.pathmorphing,matrix,angles,quotes}

\tikzstyle{thickline} = [line width=1.8pt]
\tikzstyle{gl} = [draw, gray]
\tikzstyle{pl} = [draw, orange]
\tikzstyle{ml} = [draw, blue]
\tikzstyle{sett} = [draw, thick, purple]
\tikzstyle{ele} = [draw, thick, green!70!black]

\makeatletter
\newcommand{\gettikzxy}[3]{%
  \tikz@scan@one@point\pgfutil@firstofone#1\relax
  \edef#2{\the\pgf@x}%
  \edef#3{\the\pgf@y}%
}
\makeatother

\newcommand{\settc}[1]{{\color{purple}#1}}
\newcommand{\elec}[1]{{\color{green!60!black}#1}}

\usepackage{subcaption}

\crefname{table}{Table}{Tables}
\crefname{figure}{Figure}{Figures}
\crefname{theorem}{Theorem}{Theorems}
\crefname{definition}{Definition}{Definitions}
\crefname{corollary}{Corollary}{Corollaries}
\crefname{observation}{Observation}{Observations}
\crefname{lemma}{Lemma}{Lemmas}
\crefname{example}{Example}{Examples}
\crefname{reduction}{Reduction}{Reductions}
\crefname{construction}{Construction}{Constructions}
\crefname{subsection}{Subsection}{Subsections}
\crefname{section}{Section}{Sections}
\crefname{proposition}{Proposition}{Propositions}
\crefname{algorithm}{Algorithm}{Algorithms}
\crefname{claim}{Claim}{Claims}

\tikzstyle{alter} = [draw, circle, minimum size=4ex, inner sep=1pt, text centered, align=center]

\tikzstyle{nn} = [draw, circle, inner sep=.7pt,fill=black]
\tikzstyle{sss} = [draw=red!60!black, circle, inner sep=1.2pt,fill=red!60!black]
\tikzstyle{nnn} = [draw=green!60!black, circle, inner sep=1.2pt,fill=green!60!black]
\tikzstyle{dnode} = [nn, inner sep=0.5pt, rectangle, blue, fill=blue]

\newcommand{\decprob}[3]{%
  \begin{center}%
    \begin{minipage}{0.9\linewidth}%
      \textsc{#1}\\
      \textbf{Input:} #2\\
      \textbf{Question:} #3
    \end{minipage}%
  \end{center}%
}

\newcommand{\xctg}{\textsc{Exact Cover by $3$ Sets}}
\newcommand{\xct}{\textsc{Planar and Cubic Exact Cover by $3$ Sets}}
\newcommand{\xcts}{PC-X3C}

\newcommand{\dsr}{\textsc{$\di$-SR}}

\newcommand{\dm}[1][]{{\ifthenelse{\equal{#1}{}}{$\di$}{$#1$}-matching}\xspace}
\newcommand{\dsm}[1][]{{\ifthenelse{\equal{#1}{}}{$\di$}{$#1$}-stable matching}\xspace}

\newcommand{\di}{\ensuremath{\mathsf{d}}}
\newcommand{\roomsize}{\ensuremath{n}}

\newcommand{\gthreesm}[1]{\textsc{Euclidean 3-Dimensional Stable Roommates}}
\newcommand{\gdsm}{\textsc{Euclid-3DSR}}

\newcommand{\Edsr}[1][]{\normalfont {\textsc{Euclidean} \ifthenelse{\equal{#1}{}}{$\di$}{$#1$}-\textsc{Dimensional Stable Roommates}}}
\newcommand{\edsr}[1][]{\normalfont {\textsc{Euclid}-\ifthenelse{\equal{#1}{}}{$\di$}{$#1$}-\textsc{SR}}}

\newcommand{\hs}{\ensuremath{\kappa}}

\newcommand{\sr}[1][]{\normalfont {\ifthenelse{\equal{#1}{}}{$\di$}{$#1$}-\textsc{SR}}}

\newcommand{\myemph}[1]{{\color{green!40!black}\emph{#1}}}

\newcommand{\newH}[1]{{\color{red}#1}}
\renewcommand{\newH}[1]{#1}

\newcommand{\drawphi}{
  \pic ["\scriptsize $\theta$", angle eccentricity=1.5, draw, angle radius=1.5ex] {angle=n1--v1--v2};
}
\newcommand{\drawab}{
  \pic ["\scriptsize $\quad~~ \alpha$/$\beta$", angle eccentricity=1.5, draw, angle radius=1.3ex] {angle=n1--v1--v2};
}
\newcommand{\dpentN}[6]{
  \begin{scope}[rotate=#4]
  \def\degr{10}
  \def\xx{1}
  \def\yy{1}
  \def\y{1}
  \def\nd{10}
  \def\bb{#2*#2}
  \def\aa{#1*#1}
  \def\cc{#3*#3}
  \def\bet{0}
  \def\delt{0}
  \def\alp{0}
  
  \pgfmathsetmacro\xx{#1/sin(36)/2}

  \node[nn,draw=red,fill=red] at (0,0) (n1) {};

  \node[nn] at (-#2,0) (v1) {};

  \pgfmathsetmacro\degr{180-acos((\bb+\cc-\aa)/(2*#2*#3))} 

  \node[nn] at (\degr:#3) (v2) {};
  
  \pgfmathsetmacro\degr{54+acos((\bb + \aa - \cc) / (2*#1*#2) )}
  \pgfmathsetmacro\yy{\bb+\xx*\xx-2*#2*\xx*cos(\degr)}
      
  \pgfmathsetmacro\y{sqrt(\yy)}

  \pgfmathsetmacro\bet{180-acos((\yy+\bb-\xx*\xx)/(2*\y*#2))}

  \pgfmathsetmacro\alp{acos((\yy+\xx*\xx-\bb)/(2*\y*\xx))}

  \pgfmathsetmacro\delt{\degr-108}
  
  \node at (\bet:\y) (cet) {};
  \begin{pgfonlayer}{background}
    \draw[gray!50,dashed] (n1) circle (#2);
  \end{pgfonlayer}

   \begin{scope}[shift={(cet)}]
     \foreach \i in {3,4,5} {
       \node[nn] at (\delt+\i*72-144:\xx) (v\i) {};
       \node[nn] at (\delt+\i*72-144+\alp:\y) (n\i) {};
     }
     \node[nn] at (\delt+2*72-144+\alp:\y) (n2) {};
   \end{scope}

  \foreach \i / \j in {1/2,2/3,3/4,4/5,5/1} {
    \draw[green!80!black, dashed] (v\i) -- (v\j);
    \draw[pink, thick] (v\i) -- (n\i);
    \draw (n\i) -- (v\j);
  }
\end{scope}
}

\newcommand{\dpentevenN}[6]{
  \begin{scope}[rotate=#4]
  \def\degr{10}
  \def\xx{1}
  \def\yy{1}
  \def\y{1}
  \def\nd{10}
  \def\bb{#2*#2}
  \def\aa{#1*#1}
  \def\cc{#3*#3}
  \def\bet{0}
  \def\delt{0}
  \def\alp{0}
  
  \pgfmathsetmacro\xx{#1/sin(36)/2}

  \node[nn,draw=red,fill=red] at (0,0) (n1) {};
  \node[nn,draw=red,fill=red] at (-0.25,-0.5) (n11) {};

  \node[nn] at (-#2,0) (v1) {};

  \pgfmathsetmacro\degr{180-acos((\bb+\cc-\aa)/(2*#2*#3))} 

  \node[nn] at (\degr:#3) (v2) {};
  
  \pgfmathsetmacro\degr{54+acos((\bb + \aa - \cc) / (2*#1*#2) )}
  \pgfmathsetmacro\yy{\bb+\xx*\xx-2*#2*\xx*cos(\degr)}
      
  \pgfmathsetmacro\y{sqrt(\yy)}

  \pgfmathsetmacro\bet{180-acos((\yy+\bb-\xx*\xx)/(2*\y*#2))}

  \pgfmathsetmacro\alp{acos((\yy+\xx*\xx-\bb)/(2*\y*\xx))}

  \pgfmathsetmacro\delt{\degr-108}
  
  \node at (\bet:\y) (cet) {};
  \begin{pgfonlayer}{background}
    \draw[gray!50,dashed] (n1) circle (#2);
  \end{pgfonlayer}

   \begin{scope}[shift={(cet)}]
     \foreach \i in {3,4,5} {
       \node[nn] at (\delt+\i*72-144:\xx) (v\i) {};
       \node[nn] at (\delt+\i*72-144+\alp:\y) (n\i) {};
       \node[nn] at (\delt+\i*72-144+\alp-3:\y) (n\i\i) {};
     }
     \node[nn] at (\delt+2*72-144+\alp:\y) (n2) {};
     \node[nn] at (\delt+2*72-144+\alp-3:\y) (n22) {};
   \end{scope}

  \foreach \i / \j/\p/\q in {1/2/below/above,2/3/right/left,3/4/above/left,4/5/above/below,5/1/left/right} {
    \draw[green!80!black, dashed] (v\i) -- (v\j);
    \draw[pink, thick] (v\i) -- (n\i\i);
    \draw[blue!40, thick] (v\i) -- (n\i);
    \draw (n\i) --  (v\j);
    \draw[orange!40, thick] (v\j) -- (n\i\i);
  }
\end{scope}
}

\newcommand{\dpentevennames}[2]{
  \def\xx{1}
  \pgfmathsetmacro\xx{1.5+ #1/sin(36)/2}
  \begin{scope}[rotate=#2]
   \draw[green!80!black, dashed] (v1) -- node[fill=white,text=green!50!black, inner sep=1pt] {\footnotesize $a$} (v2);
    \draw[pink, thick] (v1) -- node[below ,text=pink!80!black,inner sep=1pt] {\footnotesize $b$} (n11);
    \draw[blue!40, thick] (v1) -- (n1);
    \draw[orange!40, thick] (v2) -- (n11);
    \draw (n1) -- node[above, inner sep=2pt] {\footnotesize $c$} (v2);

  \foreach \i / \j/\p/\q in {2/3/right/left,3/4/above/left,4/5/above/below,5/1/left/right} {
    \draw[green!80!black, dashed] (v\i) -- node[fill=white,text=green!50!black, inner sep=1pt] {\footnotesize $a$} (v\j);
    \draw[pink, thick] (v\i) -- node[\p ,text=pink!80!black,inner sep=1pt] {\footnotesize $b$} (n\i\i);
    \draw[blue!40, thick] (v\i) -- node[\q ,fill=white, text=blue!60!black,inner sep=0pt] {\footnotesize $b'$} (n\i);
    \draw (n\i) -- node[pos=0.22,\q, inner sep=2pt] {\footnotesize $c$} (v\j);
    \draw[orange!40, thick] (v\j) -- node[pos=0.22,\p ,fill=white, text=orange!40!black, inner sep=0pt] {\footnotesize $c'$}(n\i\i);
  }
    
  \end{scope}
}

\newcommand{\dpentnames}[2]{
  \def\xx{1}
  \pgfmathsetmacro\xx{1.5+ #1/sin(36)/2}
  \begin{scope}[rotate=#2]
    \foreach \i / \j in {1/2,2/3,3/4,4/5,5/1} {
      \path (v\i) -- node[fill=white,text=pink!40!black,inner sep=1pt] {\footnotesize $b$} (n\i);
      \path (n\i) -- node[fill=white,inner sep=1pt] {\footnotesize $c$} (v\j);
      \path (v\i)  -- node[fill=white,text=green!50!black, inner sep=1pt] {\footnotesize $a$} (v\j);
    }
    \path[draw] (v3)  -- node[fill=white,inner sep=1pt] {\footnotesize $\ell$} (v5);  
    
  \end{scope}
}

\newcommand{\dnames}[2]{
  \def\xx{1}
  \pgfmathsetmacro\xx{1.5+ #1/sin(36)/2}
  \begin{scope}[rotate=#2]
    \foreach \i / \n / \p / \q in {1/0/below/2, 2/1/right/-1, 3/2/above/0, 4/3/left/0, 5/4/below/-2} {
      \node[\p = 0pt of v\i, xshift=\q] {\footnotesize $\n$};
    }
    \foreach \i / \n / \p / \q in {1/5/above/2,2/6/right/-1, 3/7/left/1, 4/8/left/1, 5/9/below/0} {
      \node[\p = 0pt of n\i, xshift=\q] {\footnotesize $\n$};
    }
  
  \end{scope}
}

\newcommand{\dnamesn}[2]{
  \def\xx{1}
  \pgfmathsetmacro\xx{1.5+ #1/sin(36)/2}
  \begin{scope}[rotate=#2]
    \foreach \i / \n / \p / \q / \r in {1/1/below/4/0, 2/0/right/-1/0, 3/4/above/0/0, 4/3/above/-7/-5, 5/2/below/-4/-2} {
      \node[\p = \r pt of v\i, xshift=\q] {\footnotesize $X_\n$};
    }
    \foreach \i / \n / \p / \q in {1/0/above/2,2/4/right/-1, 3/3/left/1, 4/2/left/1, 5/1/right/0} {
      \node[\p = 0pt of n\i, xshift=\q] {\footnotesize $\n$};
    }
  
  \end{scope}
}

\newcommand{\dnamesneven}[2]{
  \def\xx{1}
  \pgfmathsetmacro\xx{1.5+ #1/sin(36)/2}
  \begin{scope}[rotate=#2]
    \foreach \i / \n / \p / \q / \r in {1/1/below/4/0, 2/0/right/-1/0, 3/4/above/0/0, 4/3/above/-7/-5, 5/2/below/-4/-2} {
      \node[\p = \r pt of v\i, xshift=\q] {\footnotesize $X_\n$};
    }
    \foreach \i / \n / \p / \q in {1/0/above/2,2/8/left/-1, 3/6/left/1, 4/4/below/1, 5/2/right/0, 1/1/below/2,2/9/right/-1, 3/7/right/1, 4/5/above/1, 5/3/left/0} {
      \node[\p = 0pt of n\i, xshift=\q] {\footnotesize $\n$};
    }

  \end{scope}
}

\usepackage{etoolbox} %

\newcommand{\appsymb}{$\star$}
\newcommand{\appref}[1]{{\hyperref[#1]{\appsymb}}}

 \newcommand{\appendixproofremainPart}[5]{
    \gappto{\appendixProofText}{
  \subsection{Remaining proof of #2 of \cref{#1} for #3}\label{proof:#1}
 \noindent \emph{We continue with the proof of \cref{#1} on page \pageref{#4}:  }\\
  #5 
   }
 }

\newcommand{\appendixproof}[2]{%
 \gappto{\appendixProofText}{
  \subsection{Proof of \cref{#1}}\label{proof:#1}
  #2
   }
}

\newcommand{\appendixsection}[1]{%
  \gappto{\appendixProofText}{
    \section{Additional Material for~\cref{#1}}
    \label{appsec:#1}
  }
}

\newcommand{\dist}{\ensuremath{\delta}}
\newcommand{\distP}[1]{\ensuremath{\delta(#1,\Pi(#1))}}

\newcommand{\enn}{\ensuremath{\hat{n}}}

\newcommand{\hide}[1]{}

\newcommand{\setind}{\ensuremath{r}}

\EventEditors{Shiri Chechik, Gonzalo Navarro, Eva Rotenberg, and Grzegorz Herman}
\EventNoEds{4}
\EventLongTitle{30th Annual European Symposium on Algorithms (ESA 2022)}
\EventShortTitle{ESA 2022}
\EventAcronym{ESA}
\EventYear{2022}
\EventDate{September 5--9, 2022}
\EventLocation{Berlin/Potsdam, Germany}
\EventLogo{}
\SeriesVolume{244}
\ArticleNo{32}

\begin{document}

\maketitle

\begin{abstract}
  We investigate the \Edsr{} problem, which asks whether a given set~$V$ of $\di\cdot \roomsize$ points from
  the 2-dimensional Euclidean space can be partitioned into $\roomsize$ disjoint (unordered) subsets~$\Pi=\{V_1,\ldots,V_{\roomsize}\}$ with $|V_i|=\di$ for each $V_i\in \Pi$ such that $\Pi$ is \myemph{stable}. 
  Here, \myemph{stability} means that no point subset~$W\subseteq V$ is blocking~$\Pi$,
  and $W$ is said to be \myemph{blocking}~$\Pi$ if $|W|=\di$ such that
  $\sum_{w'\in W}\dist(w,w') < \sum_{v\in \Pi(w)}\dist(w,v)$ holds for each point~$w\in W$, 
  where $\Pi(w)$ denotes the subset~$V_i\in \Pi$ which contains~$w$
  and $\dist(a,b)$ 
  denotes the Euclidean distance between points~$a$ and $b$.
  Complementing the existing known polynomial-time result for $\di=2$, %
  we show that such polynomial-time algorithms cannot exist for any fixed number~$\di \ge 3$ unless P{}$=${}NP.
  Our result for $\di=3$ answers a decade-long open question in the theory of Stable Matching and Hedonic Games~\cite{iwama2007stable,arkin2009geometric,DeinekoWoeginger2013,Woeginger2013,Manlove2013}.
\end{abstract}

\clearpage
\section{Introduction}

We study the computational complexity of a geometric and multi-dimensional variant of the classical stable matching problem, called \Edsr{} (\edsr).
This problem is to decide whether a given set~$V$ of $\di \cdot \roomsize$ agents, each represented by a point in the two-dimensional Euclidean space~$\mathcal{R}^2$,
has a \myemph{$\di$-dimensional stable matching} (in short, \myemph{\dsm}).
Here, each agent~$x\in V$ has a preference list over all (unordered) size-$\di$ agent sets containing~$x$ which is derived from the Euclidean distances between the points.
More precisely, agent~$x$ \myemph{prefers} subset~$S$ %
to subset~$T$ %
if the sum of Euclidean distances from~$x$ to $S$ is smaller than the sum of the distances to~$T$. %
We call preferences over subsets of agents which are based on the sum of Euclidean distances \myemph{Euclidean preferences}.
A \myemph{$\di$-dimensional matching} is a partition of $V$ into $\roomsize$ disjoint agent subsets~$\Pi=\{V_1,\ldots,V_{\roomsize}\}$ with $|V_i|=\di$ for all~$i\in \{1,\ldots,\roomsize\}$.
In this way, each agent~$v\in V$ is assigned to a subset in~$\Pi$.
An agent subset~$V'$ is \myemph{blocking} the {$\di$-dimensional matching}~$\Pi$ if $|V'|=\di$ and each agent in~$V'$ prefers $V'$ to its ``assigned'' agent subset in~$\Pi$.
A \myemph{\dsm{s}} is a \dm{s} that is not blocked by a subset of agents of size $\di$.

When allowing agents to have arbitrary preferences, we arrive at the \textsc{$\di$-dimensional Stable Roommates} (\textsc{$\di$-SR}) problem with \textsc{$2$-SR} being equivalent to the classical \textsc{Stable Roommates} problem~\cite{GaleShapley1962,Irving1985}.
It is well-known that \emph{not} every instance of \textsc{Stable Roommates} admits a \dsm[2], but deciding whether there exists one is polynomial-time solvable~\cite{Irving1985}.
Fortunately, if we restrict the preferences to be Euclidean, then a \dsm[2] always exists and it can be found in polynomial time: Iteratively pick two remaining agents who are closest to each other and match them~\cite{arkin2009geometric}.
One may be tempted to apply this greedy approach to the case when~$\di=3$.
However, this would only work if it can find and match a triple of agents in each step such that this triple is the most preferred one of all three.
Since such a ``most-preferred'' triple may not always exist, the prospects become less clear.
Indeed, Arkin et al.~\cite{arkin2009geometric} showed that not every instance of \edsr[3] admits a \dsm[3].
To the best of our knowledge, nothing about the existence of \edsr{} is known for any fixed~$\di\ge 4$.
In particular, the no instance by Arkin et al.\ will not work for any fixed~$\di \ge 4$.
Arkin et al.\ left open the computational complexity of finding a \dsm[3]. %
The same question has been repeatedly asked since then~\cite{iwama2007stable,Manlove2013,DeinekoWoeginger2013,Woeginger2013,BredHeeKnoNie2020multidimensional}.
\newH{Nevertheless, \textsc{$\di$-SR} (i.e., for general preferences) has been known to be NP-complete for $\di=3$.
  Hence, it is of particular importance to search for natural restricted subcases, e.g., under Euclidean preferences, which may allow for efficient algorithms.
}

\subparagraph{Our contribution.}
In this work, we aim at settling the computational complexity of \edsr{} for all fixed $\di \ge 3$.
  Arkin et al.~\cite{arkin2009geometric} showed that there is always a 3-dimensional matching which is approximately stable, which sparks hope for a polynomial-time algorithm for $\di=3$.
  We destroy such hope by showing that \edsr[3] is NP-hard.
  We achieve this by reducing from an NP-complete planar variant of the \xctg{} problem,
  where we make use of a novel chain gadget (see the orange and blue parts in \cref{fig:edge-gadget}) and a star gadget (see \cref{fig:star}) which is adapted from the no-instance of Arkin et al.
  See the idea part in \cref{sec:3d} for more details.
 
  The same construction does not work for $\di \ge 4$ since a no-instance for \edsr[3] does not remain a no-instance for \edsr[4].
  However, we manage to derive two extended star structures, one for odd~$\di$ and the other for even~$\di$ (see the right and left figures of \cref{fig:pentagon_n-DSM}, respectively), 
  adapt the remaining component gadgets to show hardness for all fixed~$\di\ge 4$.

Together, we show the following. 
\begin{theorem}\label{thm:main}
  \edsr{} is NP-complete for every fixed~$\di \ge 3$.
\end{theorem}

\hide{
Cechl{\'a}rov{\'a} and Romero-Medina\cite{Stability2001} investigated hedonic games where every player ranks his coalitions according to the most or least attractive member of the coalition.  We use this to derive the preference of the agents which are called max preferences and min-preferences, respectively. Each agent~$x\in V$ has a preference list over all (unordered) triples containing~$x$ which is derived from the maximum (resp. minimum) Euclidean distance from the points.
More precisely, agent~$x$ \myemph{prefers} triple~$S=\{x,s_1,s_2\}$ to triple~$T=\{x,t_1,t_2\}$ if the maximum (resp. minimum) of
 Euclidean distances from~$x$ to $S$ is smaller than the maximum(resp. minimum)  of the distances to~$T$, i.e., $\max\{\dist(x,s_1),\dist(x,s_2)\} < \max\{\dist(x,t_1),\dist(x,t_2)\}$ (resp. $\min\{\dist(x,s_1),\dist(x,s_2)\} < \min\{\dist(x,t_1),\dist(x,t_2)\}$), where $\dist(a,b)$ denotes the Euclidean distance between agent~$a$ and $b$. We investigate the \gdsm{} problem under max-preferences and min-preferences. The problem of finding a core stable solution in presence of ties is known to be NP-complete under max (resp. min)-preferences. We show that \gdsm, a much restricted version of the {\sc Core Stability} problem, remain NP-complete for max-preferences, that is, we show the following.
 
}

\subparagraph{Related work.}

Knuth~\cite{knuth1976mariages} proposed to generalize the well-known \textsc{Stable Marriage} problem (a bipartite restriction of the \textsc{Stable Roommates} problem) to the 3-dimensional case. There are many such generalized variants in the literature, including the NP-complete \sr[3] problem~\cite{iwama2007stable}. 
Huang~\cite{huang2007two} strengthen the result by showing that \sr[3] remains NP-hard even for \emph{additive} preferences.
Herein, each agent~$x\in V$ has cardinal preferences~$\mu_x\colon V\setminus \{x\} \to \mathds{R}$ over all other agents such that $x$ \emph{prefers} $\{x,s_1,s_2\}$ to $\{x,t_1,t_2\}$ if and only if $\mu_x(s_1)+\mu_x(s_2) > \mu_x(t_1)+\mu_x(t_2)$.
Deineko and Woeginger~\cite{DeinekoWoeginger2013} strengthen the result of Huang by showing that
\sr[3] remains NP-hard even for metric preferences: $\mu_x(y)=\mu_y(x) \ge 0$ and $\mu_x(y)+\mu_y(z) \le \mu_{x}(z)$ such that $x$ prefers $\{x,s_1,s_2\}$ to $\{x,t_1,t_2\}$ if and only if $\mu_x(s_1)+\mu_x(s_2) < \mu_x(t_1)+\mu_x(t_2)$.
It is straightforward to see that Euclidean preferences are metric preferences and metric preferences are additive.
 We thus strengthen the results of Deineko and Woeginger, and Huang,
 by showing that the hardness remains even for Euclidean preferences.
 Recently, McKay and Manlove~\cite{mckay2021threedimensional} strengthen the result of Huang~\cite{huang2007two} by showing that the NP-hardness remains even if the cardinal preferences are binary, i.e., $\mu_x(y) \in \{0,1\}$ for all other agents~$y$.
This result is not comparable to ours since \newH{binary preferences and Euclidean preferences are not comparable.}
\newH{They also show that \sr[3] becomes polynomial-time solvable when the preferences are binary and symmetric.
}

Multi-dimensional stable matchings are equivalent to the so-called \myemph{fixed-size stable cores} in hedonic games~\cite{DG1980Hedonic}, where each coalition (i.e., a non-empty subset of agents) in the core must have the same size, and stability only needs to be guaranteed for any other coalition of the same size.\footnote{A stable core is a \emph{partition}~$\Pi$ of the agents into disjoint coalitions such that no subset of agents would block the partition~$\Pi$ by forming its own new coalition.}
Hence, our NP-hardness result also transfers to the case of finding a fixed-size stable core in the scenario where the agents in the hedonic game have Euclidean preferences. 
Hedonic games have been studied under graphical preference models~\cite{DBHSGraphHedonic,OBISY2017}, where there is an underlying social network (a directed graph) such that agents correspond to the vertices in the graph.
The general idea is to assume that agents prefer to be with their own out-neighbors more than non-out-neighbors. %
The Euclidean preference model is related to the graphical preference model where the underlying graph is planar.
However, the Euclidean model is more fine-grained and assumes that the intensity of the preferences also depends on the distance of the agents.
Notably, under the graphical model, a stable core always exists and it can be found in linear time~\cite{DBHSGraphHedonic}, but verifying whether a given partition is stable is NP-hard~\cite{CheCsaRoySim2022Verif}.
Hedonic games with fixed-size coalitions have been studied for other solution concepts such as strategy-proofness~\cite{WV2015teamformation}, Pareto optimality~\cite{CFH2019}, and exchange stability~\cite{BMM2022FixedSizeHedon}.

Other generalized variants include the study of \dsm[3] with cyclic preferences~\cite{eriksson2006three,biro2010three, %
  lam2019existence}, with preferences over individuals~\cite{iwama2007stable},
and the study of the higher-dimensional case~\cite{BredHeeKnoNie2020multidimensional} and of other restricted preference domains~\cite{BreCheFinNie2020-spscSM-jaamas}.
We refer to the textbook by Manlove~\cite{Manlove2013} for more references.

\subparagraph{Paper outline.}
In \cref{sec:prelim}, besides introducing necessary concepts and notations used throughout the paper,
we describe a crucial star-structured instance of \edsr[3] (see \cref{ex:star}), which serves as a tool of our NP-hardness reduction.
The proof of \cref{thm:main} is divided into two sections:
In \cref{sec:3d}, we consider the case of $\di=3$ and show-case in detail how to combine the star-structured instance with two new gadgets, one for the local replacement and one for the enforcement, to obtain NP-hardness.
In \cref{sec:multi}, we show how to carefully adapt the star-structured instance (which only works for $\di = 3$) and modify the reduction to show hardness for any fixed~$\di \ge 4$.
We conclude in \cref{sec:conclude}.
Due to space constraints, some figures, examples, and (part of) the proofs for results marked by \appsymb\ are deferred to the appendix. %

\section{Preliminaries}\label{sec:prelim}

Given a non-negative integer~$t$, we use ``$[t]$'' (without any prefix) to denote the set~$\{1,\ldots,t\}$.
Throughout the paper, if not stated explicitly, we assume that $\varepsilon$  and $\varepsilon_\di$ are small fractional values with $0 < \varepsilon < 0.001$ and $0< \varepsilon_{\di} < \frac{1}{1000\di}$, where $\di \ge 3$. By ``close to zero'' we mean a value which is smaller than $\varepsilon$ and $\varepsilon_{\di}$.

For each fixed integer~$\di\ge 2$, an instance of \Edsr{} (\edsr{}) consists of a set~$V=\{1,\ldots,\di \cdot \roomsize\}$ of $\di \cdot \roomsize$ agents and an embedding~$E\colon V\to \mathds{R}^{2}$ of the agents into 2-dimensional Euclidean space.
We call a non-empty subset~$V'\subseteq V$ of agents a \myemph{coalition}.
The preference list~\myemph{$\succeq_x$} of each agent~$x\in V$ over all possible size-$\di$ coalitions containing~$x$ is derived from the sum of the Euclidean distances from~$x$ to the coalition.
More precisely, for each two size-$\di$ coalitions~$S=\{x,a_1,\ldots, a_{\di-1}\}$ and $T=\{x,b_1,\ldots,b_{\di-1}\}$ containing~$x$ we say that $x$ \myemph{weakly prefers}~$S$ to $T$, denoted as
\myemph{$S \succeq_x T$}, if 
the following holds:
\begin{align*}
  \sum_{j\in [\di-1]}\dist(E(x), E(a_j)) \le   \sum_{j\in [\di-1]}\dist(E(x), E(b_j)), 
\end{align*}
where %
  $\dist(p,q) \coloneqq$ $\sqrt{(p[1]-q[1])^2+(p[2]-q[2])^2}$.
We use $S \succ_x T$ (i.e., $x$ preferring $S$ to~$T$) and $S \sim_x T$ (i.e., $x$ indifferent between $S$ and $T$) to refer to the asymmetric and symmetric part of~$\succeq_x$,
respectively.
To ease notation, for an agent~$x$ and a preference list~$\mathcal{L}$ over a subset~$\mathcal{F}$ of size-$\di$ coalitions, we use \myemph{$\mathcal{L} \succ_x \cdots$} to indicate that agent~$x$ prefers every size-$\di$ coalition in~$\mathcal{F}$ over every size-$\di$ coalition not in~$\mathcal{F}$ and her preferences over~$\mathcal{F}$ are according to~$\mathcal{L}$.
Further, we use the agent and her embedded points interchangeably,
and the distance between two agents means the distance between their embedded points.
For each agent~$x$ and each coalition~$S\subseteq V$, we use \myemph{$\dist(x,S)$} to refer to the sum of Euclidean distances from $x$ to each member in~$S$: $\dist(x,S)=\sum_{y\in S} \dist(x,y)$.

See the introduction for the definition of \myemph{\dm{s}},
\myemph{blocking coalitions}, and \myemph{\dsm{s}}.
Given a \dm{}~$\Pi$ and an agent~$x\in V$, let \myemph{$\Pi(x)$} denote the coalition that contains~$x$.
The problem studied in this paper is defined as follows:
\decprob{\edsr}
{An agent set~$V=\{1, \ldots, \di\cdot \roomsize\}$ and an embedding~$E\colon V\to \mathds{R}^2$.}
{Is there a \dsm{}?}
Note that since stability for each fixed $\di$ can be checked in polynomial time,
\edsr{} is contained in NP for every fixed $\di$.

Not every \edsr[3] instance admits a \dsm[3].
Arkin et al.~\cite{arkin2009geometric} provided a star-structured instance which does not. %
In \cref{ex:star}, we describe an \emph{adapted} variant of their instance, which is a decisive component of our hardness reduction.

\begin{figure}
    \centering
  \begin{tikzpicture}[scale=.13]

  \begin{scope}%
    \dpentN{6.6}{10.1}{10.2}{0}{10.1}{10.2}
    \dpentnames{6.6}{0}

    \dnames{6.6}{0}

    \drawphi
  \end{scope}
  \begin{scope}
    \node[dnode] at ($(n1)+(-30:9.95)$) (x) {};
    \node[dnode] at ($(n1)+(-34:9.95)$) (y) {};
    \node[above=0pt of x] {\footnotesize 10};
    \node[below=0pt of y] {\footnotesize 11};
    \draw (n1) -- node[above, inner sep=1pt, text=pink!40!black, yshift=1] {\small $~<\! b$} (x);
    \draw (n1) -- (y);
    \draw (x) -- (y);
  \end{scope}
\end{tikzpicture}
\caption{A star-structured instance adapted from Arkin et al.~\cite{arkin2009geometric}; see  \cref{ex:star}. We use different colors to highlight the distances between the points. For instance, the smallest distance between any two points is $a$ (highlighted in green). We also draw a dashed circle of radius~$b$, centered at point~$5$ to indicate that points both $10$ and $11$ are with distance smaller than~$b$ to $5$.}\label{fig:star}
\end{figure}
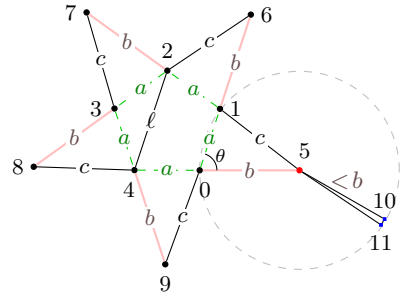

\begin{example}\label{ex:star}
  Consider an instance which contains \emph{at least} $12$ agents called~$W=\{0,\ldots, 11\}$ where the $12$ agents are embedded as given in \cref{fig:star}. %
  In the embedding of $W\setminus \{10,11\}$, the five inner-most points, namely~$0$ to $4$, form a regular pentagon with edge length~$a$.
  For each $i\in \{0,\ldots, 4\}$, the three points~$i$, $i+1 \bmod 5$, and $i+5$ form a triangle with side lengths~$a, b, c$ such that $a < b < c < \ell$, where $\ell$ denotes the diagonal of the regular pentagon.
  Moreover, the angle~$\theta$ at points~$i+1 \bmod 5$, $i$, $i+5$ is at most 90 degrees. This ensures that the distance between points~$(i+1 \bmod 5) + 5$ and $i$ is strictly larger than $\ell$ (we will use this fact later).
  Except for point~$5$ (marked in red), the closest neighbor of each point~$i+5$ is $i$, followed by $i+1 \bmod 5$. %
  Point~$5$'s two closest neighbors are points~$10$ and $11$ with $a < \dist(5, 10) < b$ and $a < \dist(5, 11) < b$,
  followed by points~$0$ and~$1$.
  The distance between $10$ and $11$ is close to zero, with the intention to ensure that every
  \dsm[3] must match them together.
  The distance from $10$ (resp.\ $11$) to any agent in~$W\setminus \{5,10,11\}$ is larger than the diagonal length~$\ell$ while the distance from~$10$ (resp.\ $11$) to any agent not in~$W$ is larger than $\dist(5,10)-\varepsilon$. %
  Finally, the distance between any agent from~$W\setminus \{10,11\}$ to any agent \emph{not} from~$W\setminus \{10,11\}$ is strictly larger than~$\ell$.
  To specify the embedding of the agents from~$W$, we use the polar coordinate system. %
  We first fix the embeddings of~$5,10,11$ to ensure the distances between them are as stated above.
  Then, we fix points~$0$ and $1$ and the centroid of the regular pentagon to
  ensure the distances satisfy~$a<b<c<\ell$, and the angle~$\theta$ at points~$1,0,5$ is at most 90 degrees,
  and the angle at points~$0,5,j$, $j\in \{10,11\}$, is more than 90 degrees. 
  Once these points are fixed we can determine the other points by a simple calculation.
\end{example}

The instance of Arkin et al.~\cite{arkin2009geometric} embeds the two extra points~$10$ and $11$ differently
than ours (see \cref{ex:star}).
Hence, their instance is a no-instance, while ours may be a yes-instance, provided some specific triple is matched together, formulated as follows:

\begin{lemma}\label{lem:pentagon}
  Every \dsm[3] of an instance satisfying the embedding described in \Cref{ex:star} must contain triple~$\{5,10,11\}$.
\end{lemma}

\begin{proof}

  Towards a contradiction, suppose that $\Pi$ is a \dsm[3] with $\{5,10,11\}\notin \Pi$.
  We infer that $\{10,11\}\subseteq \Pi(10)$ since otherwise $\{5,10,11\}$ is blocking~$\Pi$ due the following:
    $\dist(5,\Pi(5)) \ge \min(\dist(5,10), \dist(5,11)) + b > \dist(5,\{5,10,11\})$,
    and for each~$x\in \{10,11\}$ it holds that
    $\dist(x,\Pi(x)) \ge 2(\dist(x,5)-\varepsilon) >\dist(x,5) + \dist(10,11)$ for any $\varepsilon >0$.
    This implies that $\{10,11\}\cap \Pi(5) = \emptyset$.
  Next, we observe that there must be a triple in~$\Pi$ that contains the two agents of at least one pentagon edge as otherwise $\{2,3,7\}$ is blocking: $\dist(2,\Pi(2)) \ge b+c > a+b$, $\dist(3,\Pi(3)) \ge b+c > a+c$, and $\dist(7,\Pi(7)) \ge b+\ell> b+c$.
  Thus, at least one triple in~$\Pi$ contains the agents of some pentagon edge, say~$\{2,3\}$; the other cases are analogous.
  Let $\{2,3,x\}\in \Pi$.
  We distinguish between three subcases: %

  \begin{description}
    \item[Case 1: $x\notin \{1,4,7,8\}$.]
    Then, one can verify that $\{2,3,7\}$ is blocking; recall that every agent not in~$W\setminus \{2,3\}$ is at distance larger than $\ell$ to agent~$7$.

    \item[Case 2: $x \in \{1,7\}$.]
    Then, $\Pi(4) = \{0,4,9\}$ or $\Pi(4)=\{0,4,8\}$ since otherwise $\{3,4,8\}$ blocks~$\Pi$
    due to:
    $\dist(3,\Pi(3)) \ge a+\min(\dist(3,1), \dist(3,7)) = a+c> a+b$,
    $\dist(4,\Pi(4)) > a+c$,
    $\dist(8,\Pi(8)) > b+c$  (recall that the distance from every agent not in~$W\setminus \{3,4\}$ to agent~$8$ is larger than $\ell$).    
    However, both cases imply that 
    $\{0,1,5\}$ is blocking since
    $\dist(0,\Pi(0)) \ge a+c > a+b = \dist(0, \{0,1,5\})$,
    $\dist(1,\Pi(1)) \ge a + \ell > a+c = \dist(1, \{0,1,5\})$, and
    $\dist(5,\Pi(5)) \ge c+\ell > b+c = \dist(5, \{0,1,5\})$; recall that $\Pi(5)\cap \{10,11\} = \emptyset$.
  
    \item[Case 3: $x \in \{4,8\}$.]
    Then, $\dist(2, \Pi(2)) \ge a+\ell > a+c$.
    This implies that $\{0,1,6\} \in \Pi$ since otherwise $\{1,2,6\}$ is blocking~$\Pi$.
    However, this implies that $\{0,4,9\}$ is blocking~$\Pi$.
\end{description}
\noindent Since we have just shown that no agent~$x$ exists which is in the same triple as $2$ and $3$,
no \dsm[3] exists that does not contain $\{5,10,11\}$.
\end{proof}

\section{NP-hardness for \edsr[3]}\label{sec:3d}

In this section, we prove \cref{thm:main} for the case of~$\di=3$ by providing a polynomial reduction from the NP-complete \xct\ problem~\cite{MooreRobson2001Tiling}, which is 
an NP-complete restricted variant of the \xctg{} problem~\cite{GJ79}. %

\decprob{\xct\ (\xcts)}
{A $3n$-element set~$X=\{1,\ldots,3n\}$ and a collection~$\mathcal{S}=(S_1,\ldots,S_{m})$ of $3$-element subsets of~$X$ of cardinality~$3n$ such that each element occurs in exactly three sets and the associated graph is planar.}{Does~$\mathcal{S}$ contain an \emph{exact cover} for~$X$, i.e., a subcollection~$\mathcal{K} \subseteq \mathcal{S}$ such that each element of~$X$ occurs in exactly one member of~$\mathcal{K}$?}

\noindent Herein, given a \xcts{} instance~$I=(X,\mathcal{S})$, the associated graph of~$I$, denoted as {$G(I)$}, is a bipartite graph~\myemph{$G(I)=(U \uplus W, E)$} on two partite vertex sets~$U=\{u_i\mid i \in X\}$ and $W=\{w_j \mid S_j\in \mathcal{S}\}$ such that there exists an edge~$e=\{u_i,w_j\}\in E$ if and only if $i\in S_j$. We call the vertices in~$U$ and $W$ the \myemph{element-vertices} and the \myemph{set-vertices}, respectively.
In our reduction, we crucially utilize the fact that the associated graph $G$ of the input instance is planar and cubic, and hence by Valiant~\cite{Valiant1981} admits a specific planar embedding in~$\mathds{Z}^2$, called \myemph{orthogonal drawing}, which maps each vertex to an integer grid point and each edge to a chain of non-overlapping horizontal and vertical segments along the grid (except at the endpoints).
To simplify the description of the reduction, we %
use the following more restricted orthogonal drawing:
\begin{proposition}[\cite{Battista1998orthogonal}]\label{prop:orthogonal}
  In polynomial time, a planar graph with maximum vertex degree three can be embedded in the grid $\mathds{Z}^2$ such that its vertices are at the integer grid points and its edges are drawn using at most one horizontal and one vertical segment in the grid. 
\end{proposition}
We call the intersection point of the horizontal and vertical segments the \myemph{bending point}.

\appendixsection{sub:3-sr-construction}
\subsection{The construction}\label{sub:3-sr-construction}
\subparagraph{The idea.}

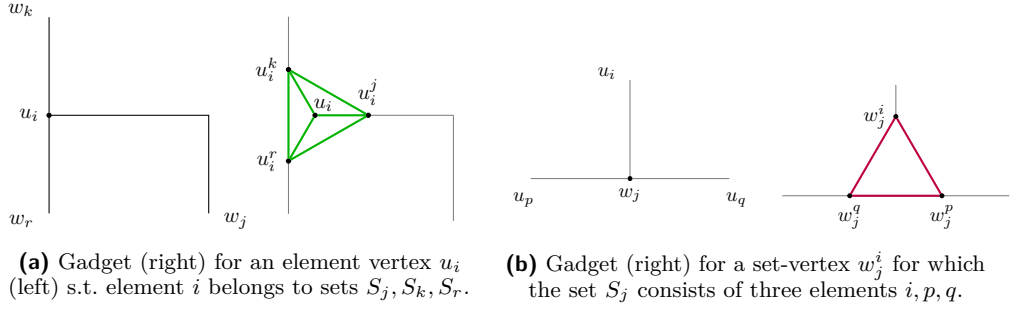
\begin{figure}[t!]
  \captionsetup[subfigure]{justification=centering}
  \centering
  \begin{subfigure}[t]{.45\textwidth}
    \begin{tikzpicture}[scale=1.4,every node/.style={scale=0.8}]
  \node[nn] at (0,0) (ui) {};
  \node[left = 0pt of ui] {$u_i$};
  \node at (1.5,-1) (s1) {};
  \node at (0,-1) (s3) {};
  \node at (0,1) (s2) {};

  \node[right=0pt of s1] {$w_j$};
  \node[left=0pt of s2] {$w_k$};
  \node[left=0pt of s3] {$w_{\setind}$};

  \path[draw] (ui) -| (s1);
  \path[draw] (ui) -- (s2);
  \path[draw] (ui) -- (s3);

  \begin{scope}[shift={(2.5,0)}]
    \node[nn] at (0:0.5) (s1) {};
    \node[nn] at (120:0.5) (s2) {};
    \node[nn] at (240:0.5) (s3) {};
    \node[nn] at (0:0) (i) {};

    \node[above=0pt of s1] {$u_i^j$};
    \node[left= 0pt of s2] {$u_i^k$};
    \node[left=0pt of s3] {$u_i^{\setind}$};

    \node[above right=-2pt and -4pt of i] {$u_i$};

    \foreach \i / \j in {i/s1,i/s2,i/s3,s1/s2,s2/s3,s3/s1} {
      \path[ele] (\i) -- (\j);
    }

    \begin{pgfonlayer}{background}
      \begin{scope}[shift={(s2)}]
        \path[gl] (0,0) -- (0,0.5);
      \end{scope}
      
      \begin{scope}[shift={(s3)}]
        \path[gl] (0,0) -- (0,-0.5);
      \end{scope}

      \begin{scope}[shift={(s1)}]
        \path[gl] (0,0) -- (.8,0);
        \path[gl] (.8,0) -- (.8,-1);
      \end{scope}
    \end{pgfonlayer}
    
  \end{scope}
\end{tikzpicture}
\caption{Gadget (right) for an element vertex~$u_i$ (left) s.t.\  element~$i$ belongs to sets~$S_j,S_k,S_{\setind}$.}\label{fig:elt-gad}
\end{subfigure}~~
\begin{subfigure}[t]{0.45\linewidth}
  \begin{tikzpicture}[scale=1.4,every node/.style={scale=0.75}]
  \node[nn] at (0,0) (ui) {};
  \node[below = 0pt of ui] {$w_j$};
  \node at (-1,0) (s1) {};
  \node at (1,0) (s3) {};
  \node at (0,1) (s2) {};

  \node[below=0pt of s1] {$u_p$};
  \node[left=0pt of s2] {$u_i$};
  \node[below=0pt of s3] {$u_q$};

  \path[gl] (ui) -- (s1);
  \path[gl] (ui) -- (s2);
  \path[gl] (ui) -- (s3);

  \begin{scope}[shift={(2:2.5)}]
    \node[nn] at (-30:0.5) (s1) {};
    \node[nn] at (90 :0.5) (s2) {};
    \node[nn] at (210:0.5) (s3) {};

    \node[below=0pt of s1] {$w_j^p$};
    \node[left= 0pt of s2] {$w_j^i$};
    \node[below =0pt of s3] {$w_j^{q}$};

    \foreach \i / \j in {s1/s2,s2/s3,s3/s1} {
      \path[sett] (\i) -- (\j);
    }
    
    \begin{pgfonlayer}{background}
      \begin{scope}[shift={(s2)}]
        \path[gl] (0,0) -- (0,0.3);
      \end{scope}

      \begin{scope}[shift={(s3)}]
        \path[gl] (0,0) -- (1.5,0);
      \end{scope}

      \begin{scope}[shift={(s1)}]
        \path[gl] (0,0) -- (-1.5,0);
      \end{scope}
    \end{pgfonlayer}
  \end{scope}
\end{tikzpicture}
\caption{Gadget (right) for a set-vertex~$w_j^i$ for which the set~$S_j$ consists of three elements~$i,p,q$.}\label{fig:set-gad}
\end{subfigure}
\caption{Element- and set-gadgets described in \cref{sub:element-set-gadgets}.}\label{fig:elem-set-gadgets}
\end{figure}

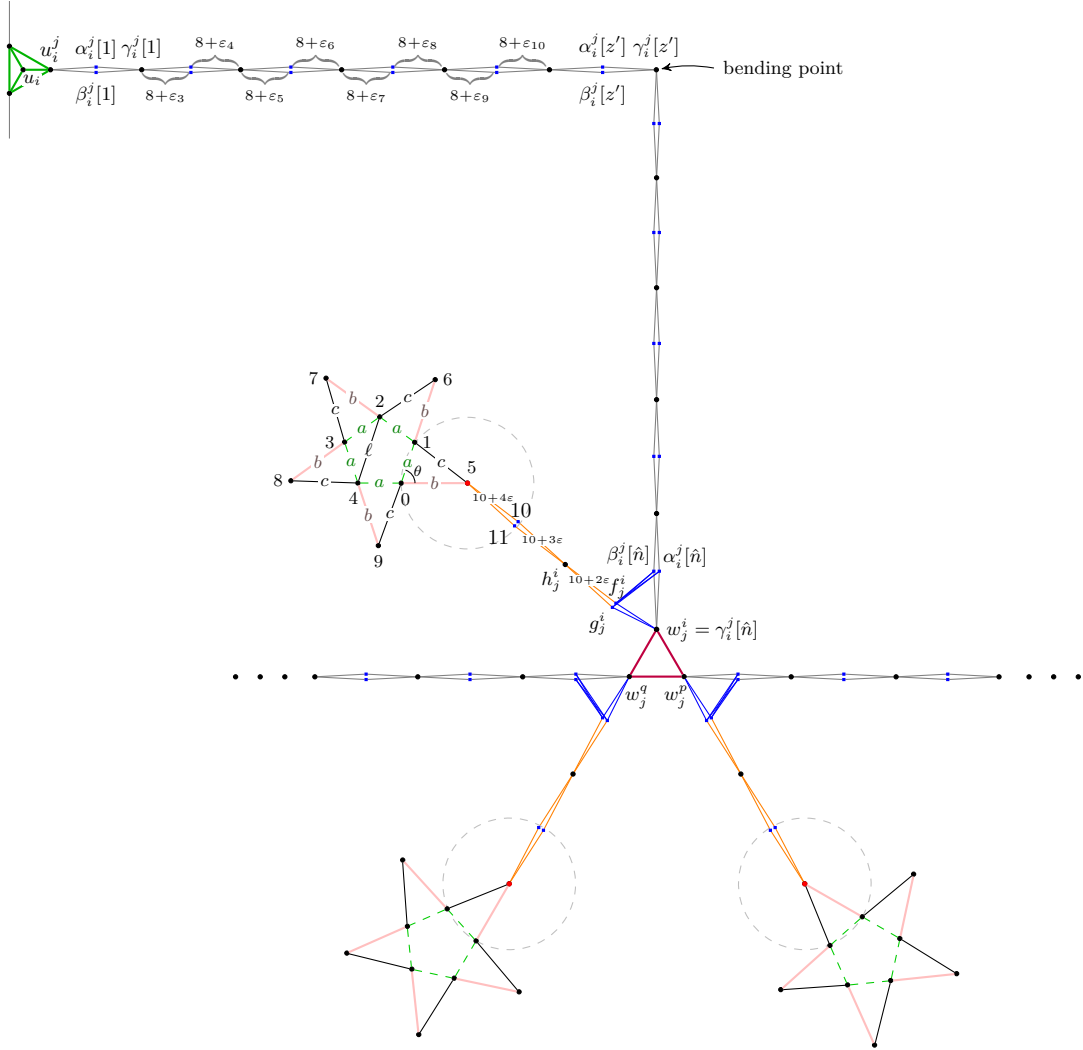
\begin{figure}[t!]
  \centering
\begin{tikzpicture}[scale=1.2, every node/.style={scale=0.8}, >=stealth']
  \def\d{.54}
  \def\f{.6}
  \def\e{.3}
  \def\k{.5}
  
  \node[nn] at (0:\e) (s1) {};
  \node[nn] at (120:\e) (s2) {};
  \node[nn] at (240:\e) (s3) {};
  \node[nn] at (0:0) (i) {};
  
  \node[above=-1pt of s1] {$u_i^j$};

  \foreach \i / \j in {i/s1,i/s2,i/s3,s1/s2,s2/s3,s3/s1} {
    \path[ele] (\i) -- (\j);
  }

  \node[below right=1pt and -1pt of i, inner sep=.2pt, fill=white] {\small $u_i$};

  \begin{pgfonlayer}{background}
  \begin{scope}[shift={(s2)}]
    \path[gl] (0,0) -- (0,0.5);
  \end{scope}
  
  \begin{scope}[shift={(s3)}]
    \path[gl] (0,0) -- (0,-0.5);
  \end{scope}
  \end{pgfonlayer}
    \coordinate (x) at (s1);
    \begin{scope}[shift={(0.8,0)}]
      \node[dnode] at (0,0.03) (f1) {};
      \node[dnode] at (0,-0.03) (f2) {};
      \node[nn] at (\k,0) (f3) {};
      \begin{pgfonlayer}{background}
        \draw[gl] (x) -- (f1) -- (f3);
        \draw[gl] (x) -- (f2) -- (f3);
      \end{pgfonlayer}
      \node[above = -0.5pt of f1] {\small $\alpha^{j}_i[1]$};
      \node[below = 0.5pt of f2] {\small $\beta^{j}_i[1]$};
      \node[above = 0.5pt of f3] {\small $\gamma^j_i[1]$};
    \end{scope}

    \coordinate (x) at (f3);
    \def\st{1.3}
    
     \foreach \i/\j/\k/\l in {1/2.01/3/4,3.03/4.06/5/6,
      5.1/6.15/7/8,7.21/8.28/9/10} {
      \begin{scope}[shift={(\st,0)}]
        \node[dnode] at (\i*\d,0.03) (f1) {};
        \node[dnode] at (\i*\d,-0.03) (f2) {};
        \node[nn] at (\j*\d,0) (f3) {};
        \begin{pgfonlayer}{background}
          \draw[gl] (x) -- (f1) -- node[text=gray, above, inner sep=0pt] (f13) {\small $\overbrace{~~~~~~~}$} (f3);
          \node[above = -0.5pt of f13] {\scriptsize $8\!+\!\varepsilon_{\l}$};
          \draw[gl] (x) -- node[text=gray, below, inner sep=0pt] (f23) {\small $\underbrace{~~~~~~~}$} (f2) -- (f3);
          \node[below = -0.5pt of f23] {\scriptsize $8\!+\!\varepsilon_{\k}$};
        \end{pgfonlayer}
      \end{scope}
      \coordinate (x) at (f3);
    }
    
         \foreach \i/\j/\k/\l in {9.36/10.45/11/12} {
      \begin{scope}[shift={(\st,0)}]
        \node[dnode] at (\i*\d,0.03) (f1) {};
        \node[dnode] at (\i*\d,-0.03) (f2) {};
        \node[nn] at (\j*\d,0) (f3) {};
        \begin{pgfonlayer}{background}
          \draw[gl] (x) -- (f1) --  (f3);
          \draw[gl] (x) --  (f2) -- (f3);
        \end{pgfonlayer}
      \end{scope}
      \coordinate (x) at (f3);
    }

  \node[above = 0.5pt of f1] {\small $\alpha^{j}_i[z']$};
  \node[below = 0.5pt of f2] {\small $\beta^{j}_i[z']$};
  \node[above = 1pt of f3] {\small $\gamma^j_i[z']$};
  \node[draw=none, right = 5ex of f3] (bendingpoint) {};
  \node[right = -6pt of bendingpoint] {\small bending point};

  \draw[->, shorten <= 1pt, shorten >= 1pt] (bendingpoint) edge[bend right = 10] (f3);

  \begin{scope}[shift={(x)}]
    \foreach \i/\j in {-1.1/-2.21,-3.33/-4.46,-5.6/-6.75,-7.91/-9.08,-10.26/-11.45} {
      \begin{scope}
        \node[dnode] at (-0.03,\i*\d) (f1) {};
        \node[dnode] at (0.03,\i*\d) (f2) {};
        \node[nn] at (0,\j*\d) (f3) {};
        \begin{pgfonlayer}{background}
          \draw[gl] (x) -- (f1) -- (f3);
          \draw[gl] (x) -- (f2) -- (f3);
        \end{pgfonlayer}
      \end{scope}
      \coordinate (x) at (f3);
    }
  \end{scope}
  \coordinate (an) at (f1);
  \coordinate (bn) at (f2);
  
  \node[above left = -1pt and  -2pt of f1] {\small $\beta^j_i[{\enn}]$};
  \node[above right = -2pt and  -2pt of f2] {\small $\alpha^j_i[{\enn}]$};
  \begin{scope}[shift={(x)}]
    \node[nn] at (x) (s1) {};
    \node[nn] at (-60:\f) (s2) {};
    \node[nn] at (240:\f) (s3) {};

    \node[right=0pt of s1] {\small $w_j^i=\gamma^j_i[{\enn}]$};
    \node[below left = 0pt and -5pt of s2] {\small $w_j^p$};
    \node[below right = 0pt and -5pt of s3]  {\small $w_j^{q}$};

    \foreach \i / \j in {s1/s2,s2/s3,s3/s1} {
      \path[sett] (\i) -- (\j);
    }
  \end{scope}

  \coordinate (x) at (s3);
  \begin{scope}[shift={(x)}]
    \foreach \i / \j in {-1.09/-2.17} {
      \begin{scope}
        \node[dnode] at (\i*\d,-0.03) (f1) {};
        \node[dnode] at (\i*\d,0.03) (f2) {};
        \node[nn] at (\j*\d,0) (f3) {};
        \begin{pgfonlayer}{background}
          \draw[gl] (x) -- (f1) -- (f3);
          \draw[gl] (x) -- (f2) -- (f3);
        \end{pgfonlayer}
      \end{scope}
      \coordinate (x) at (f3);
    }

    \coordinate (aqn) at (f1);
    \coordinate (bqn) at (f2);
    
    \foreach \i/\j in {-3.24/-4.3,-5.35/-6.39} {
      \begin{scope}
        \node[dnode] at (\i*\d,-0.03) (f1) {};
        \node[dnode] at (\i*\d,0.03) (f2) {};
        \node[nn] at (\j*\d,0) (f3) {};
        \begin{pgfonlayer}{background}
          \draw[gl] (x) -- (f1) -- (f3);
          \draw[gl] (x) -- (f2) -- (f3);
        \end{pgfonlayer}
      \end{scope}
      \coordinate (x) at (f3);
    }

    \foreach \i in {-7,-7.5,-8} {
        \node[nn] at (\i*\d,0) {};
    }
    \end{scope}    
    
    \coordinate (x) at (s2);
    \begin{scope}[shift={(x)}]
      \foreach \i / \j in {1.09/2.17} {
      \begin{scope}
        \node[dnode] at (\i*\d,-0.03) (f1) {};
        \node[dnode] at (\i*\d,0.03) (f2) {};
        \node[nn] at (\j*\d,0) (f3) {};
        \begin{pgfonlayer}{background}
          \draw[gl] (x) -- (f1) -- (f3);
          \draw[gl] (x) -- (f2) -- (f3);
        \end{pgfonlayer}
      \end{scope}
      \coordinate (x) at (f3);
    }
    
    \coordinate (apn) at (f1);
    \coordinate (bpn) at (f2);
    
   \foreach \i/\j in {3.24/4.3,5.35/6.39} {
      \begin{scope}
        \node[dnode] at (\i*\d,-0.03) (f1) {};
        \node[dnode] at (\i*\d,0.03) (f2) {};
        \node[nn] at (\j*\d,0) (f3) {};
        \begin{pgfonlayer}{background}
          \draw[gl] (x) -- (f1) -- (f3);
          \draw[gl] (x) -- (f2) -- (f3);
        \end{pgfonlayer}
      \end{scope}
      \coordinate (x) at (f3);
    }

    \foreach \i in {7,7.5,8} {
      \node[nn] at (\i*\d,0) {};
    }

    \end{scope}

    \def\f{0.7}

    \node[] at ([shift={(150:\d)}]s1) (x) {};
    
    \begin{scope}[shift={(x)},rotate=-40]
      \foreach \i in {0} {
        \begin{scope}[shift={(\i*\f,0)}]
          \node[dnode] at (0,-0.03) (f1) {};
          \node[dnode] at (0,0.03) (f2) {};
          \node[nn] at (-\f,0) (f3) {};

          \begin{pgfonlayer}{background}
            \foreach \s in {f1,f2} {
              \foreach \t in {s1,an,bn}
              {
                \draw[ml] (\s) -- (\t);
              }
            }
            \draw[pl] (f2) -- (f3);
            \draw[pl] (f1) -- node[text=black, above, fill=white, inner sep=0.5pt] {\tiny $10\!+\!2\varepsilon$} (f3);
          \end{pgfonlayer}
        \end{scope}
        \coordinate (x) at (f3);
      }
      \node[above right = 0pt and -5pt of f1] {\small $f^i_j$};
      \node[below left = 0pt and 0pt of f2] {\small $g^i_j$};
      \node[below left = -2pt and -2pt of f3] {\small $h^i_j$};
      
      \foreach \i in {-2} {
        \begin{scope}[shift={(\i*\f,0)}]
          \node[dnode] at (0,-0.03) (f1) {};
          \node[dnode] at (0,0.03) (f2) {};
          \node[nn] at (-\f,0) (f3) {};
          \begin{pgfonlayer}{background}
            \draw[pl] (x) -- (f1) -- (f3);
            \draw[pl] (x) -- node[text=black, above, fill=white, inner sep=0.5pt,yshift=-1pt] {\tiny $10\!+\!3\varepsilon$} (f2) -- node[text=black, above, fill=white, inner sep=0.5pt,yshift=-1pt] {\tiny $10\!+\!4\varepsilon$} (f3);
          \end{pgfonlayer}
        \end{scope}
        \coordinate (x) at (f3);
      }
      \node[above right = 0pt and -5pt of f1] {$10$};
      \node[below left = 0pt and 0pt of f2] {$11$};
    \end{scope}

    \coordinate (x) at (f3);
    
    \begin{scope}[scale=0.072,shift={(x)}]
      \def\xx{1}
      \def\yy{1}
      \def\degr{10}
      \def\y{1}
      \def\nd{10}
      \def\a{6.6}
      \def\b{10.1}
      \def\c{10.2}
      \def\bb{\b*\b}
      \def\aa{\a*\a}
      \def\cc{\c*\c}

      \pgfmathsetmacro\xx{\a/sin(36)/2}
      
      \pgfmathsetmacro\degr{54+acos((\bb + \aa - \cc) / (2*\a*\b) )}
      \pgfmathsetmacro\yy{\bb+\xx*\xx-2*\b*\xx*cos(\degr)}
      
      \pgfmathsetmacro\y{sqrt(\yy)}
      \pgfmathsetmacro\nd{acos((\xx*\xx+\yy-\bb)/(2*\xx*\y))}

      \begin{scope}
        \dpentN{\a}{\b}{\c}{0}{10.1}{10.2}
        \dpentnames{\a}{0}
        \dnames{\a}{0}
        \drawphi
      \end{scope}
    \end{scope}

     \node[] at ([shift={(-60:\d)}]s2) (x) {};
    
    \begin{scope}[shift={(x)},rotate=-60]
      \foreach \i in {0} {
        \begin{scope}[shift={(\i*\f,0)}]
          \node[dnode] at (0,-0.03) (f1) {};
          \node[dnode] at (0,0.03) (f2) {};
          \node[nn] at (\f,0) (f3) {};

          \begin{pgfonlayer}{background}
            \foreach \s in {f1,f2} {
              \foreach \t in {s2,apn,bpn}
              {
                \draw[ml] (\s) -- (\t);
              }
            }
            \draw[pl] (f1) -- (f3);
            \draw[pl] (f2) -- (f3);
          \end{pgfonlayer}
        \end{scope}
        \coordinate (x) at (f3);
      }
      
      \foreach \i in {2} {
        \begin{scope}[shift={(\i*\f,0)}]
          \node[dnode] at (0,-0.03) (f1) {};
          \node[dnode] at (0,0.03) (f2) {};
          \node[nn] at (\f,0) (f3) {};
          \begin{pgfonlayer}{background}
            \draw[pl] (x) -- (f1) -- (f3);
            \draw[pl] (x) -- (f2) -- (f3);
          \end{pgfonlayer}
        \end{scope}
        \coordinate (x) at (f3);
      }
    \end{scope}

    \coordinate (x) at (f3);

    \begin{scope}[scale=0.072,shift={(x)}]
      \def\xx{1}
      \def\yy{1}
      \def\degr{10}
      \def\y{1}
      \def\nd{10}
      \def\a{6.6}
      \def\b{10.1}
      \def\c{10.2}
      \def\bb{\b*\b}
      \def\aa{\a*\a}
      \def\cc{\c*\c}

      \pgfmathsetmacro\xx{\a/sin(36)/2}
      
      \pgfmathsetmacro\degr{54+acos((\bb + \aa - \cc) / (2*\a*\b) )}
      \pgfmathsetmacro\yy{\bb+\xx*\xx-2*\b*\xx*cos(\degr)}
      
      \pgfmathsetmacro\y{sqrt(\yy)}
      \pgfmathsetmacro\nd{acos((\xx*\xx+\yy-\bb)/(2*\xx*\y))}

      \begin{scope}
        \dpentN{\a}{\b}{\c}{150}{10.1}{10.2}
      \end{scope}
    \end{scope}

     \node[] at ([shift={(240:\d)}]s3) (x) {};
    
    \begin{scope}[shift={(x)},rotate=240]
      \foreach \i in {0} {
        \begin{scope}[shift={(\i*\f,0)}]
          \node[dnode] at (0,-0.03) (f1) {};
          \node[dnode] at (0,0.03) (f2) {};
          \node[nn] at (\f,0) (f3) {};

          \begin{pgfonlayer}{background}
            \foreach \s in {f1,f2} {
              \foreach \t in {s3,aqn,bqn}
              {
                \draw[ml] (\s) -- (\t);
              }
            }
            \draw[pl] (f1) -- (f3);
            \draw[pl] (f2) -- (f3);
          \end{pgfonlayer}
        \end{scope}
        \coordinate (x) at (f3);
      }
      
      \foreach \i in {2} {
        \begin{scope}[shift={(\i*\f,0)}]
          \node[dnode] at (0,-0.03) (f1) {};
          \node[dnode] at (0,0.03) (f2) {};
          \node[nn] at (\f,0) (f3) {};
          \begin{pgfonlayer}{background}
            \draw[pl] (x) -- (f1) -- (f3);
            \draw[pl] (x) -- (f2) -- (f3);
          \end{pgfonlayer}
        \end{scope}
        \coordinate (x) at (f3);
      }
    \end{scope}

    \coordinate (x) at (f3);

    \begin{scope}[scale=0.072,shift={(x)}]
      \def\xx{1}
      \def\yy{1}
      \def\degr{10}
      \def\y{1}
      \def\nd{10}
      \def\a{6.6}
      \def\b{10.1}
      \def\c{10.2}
      \def\bb{\b*\b}
      \def\aa{\a*\a}
      \def\cc{\c*\c}

      \pgfmathsetmacro\xx{\a/sin(36)/2}
      
      \pgfmathsetmacro\degr{54+acos((\bb + \aa - \cc) / (2*\a*\b) )}
      \pgfmathsetmacro\yy{\bb+\xx*\xx-2*\b*\xx*cos(\degr)}
      
      \pgfmathsetmacro\y{sqrt(\yy)}
      \pgfmathsetmacro\nd{acos((\xx*\xx+\yy-\bb)/(2*\xx*\y))}

      \begin{scope}
        \dpentN{\a}{\b}{\c}{60}{10.1}{10.2}
      \end{scope}
    \end{scope}
  \end{tikzpicture}
  \caption{Gadget for edge~$\{u_i,w_j\}$ in $G(I)$ with $S_j=\{i,p,q\}$.
    Here, the fractional values~$\varepsilon_z$ satisfy $0<\varepsilon_1 < \cdots < \varepsilon_{\enn} = 2-\varepsilon$.
    The star-gadget, adapted from Arkin et al.~\cite{arkin2009geometric}, is described in  \cref{ex:star}.
    To highlight the distances between the points in the star-gadget, we use different colors.
    For instance, the smallest distance between any two points in the star is $a$ (highlighted in green). We also draw a dashed circle of radius~$b$, centered at point~$5$ to indicate that points both $10$ and $11$ are with distance smaller than $b$ to $5$.}\label{fig:edge-gadget} %
\end{figure}

Given an instance~$I=(X,\mathcal{S})$ of \xcts{}, we first use \cref{prop:orthogonal} to embed the associated graph~$G(I)=(U\uplus W, E)$ into a $2$-dimensional grid with edges drawn using line segments of length at least $L\ge 200$, and with parallel lines at least $4L$ grid squares apart.
The idea is to replace each element-vertex~$u_i\in U$ with four agents which form a ``star'' with three close-by ``leaves'' (see \cref{fig:elt-gad}).
These leaves one-to-one correspond to the sets~$S_j$ with $i\in S_j$.
In this way, exactly one set~$S_j$ is unmatched with the center and will be chosen to the exact cover solution.
Furthermore, we replace each set-vertex~$w_j\in W$ with three agents~$w_j^i$, $i\in S_j$, which form an equilateral triangle (see \cref{fig:set-gad}).
We replace each edge in~$G(I)$ with a chain of copies of three agents, which, together with a private enforcement gadget (the star structure with a tail in \cref{fig:edge-gadget}), ensure that either all three agents~$w_j^i$
are matched in the same triple (indicating that the corresponding set is in the solution) or none of them is matched in the same triple (indicating that the corresponding set is not in the solution). {The agents in the star structure can be embedded ``far'' from other agents due to the tail.}

\subparagraph{Gadgets for the elements and the sets.}\label{sub:element-set-gadgets}
For each element-vertex~$u_i\in U$, assume that the three connecting edges in $G(I)$ are going horizontally to the right (rightward), vertically up (upward), and vertically down (downward); we can mirror the coordinate system if this is not the case.
Let $w_j, w_k, w_{\setind}$ denote the set-vertices on the endpoints of the rightward, upward, and  downward edge, respectively.
We create four element-agents, called $u_i$, $u_i^j$, $u_i^k$, and $u_i^{\setind}$.
We embed them into~$\mathds{R}^2$ in such a way that $u_i^j,u_i^k,u_i^{\setind}$ are on the segment of the rightward, upward, and downward edge, respectively,
and are of equal distance \elec{$8$} to each other.
Agent~$u_i$ is in the center of the other three agents.
See Figure~\ref{fig:elt-gad} for an illustration.

Similarly, for each set-vertex~$w_j\in W$, assume that the three connecting edges in~$G(I)$ are going rightward, leftward, and upward, connecting the element-vertices~$u_i$, $u_p$, $u_q$, respectively.
We create three set-agents, called $w_j^i$, $w_j^p$, $w_j^q$.
We embed them into~$\mathds{R}^2$ in such a way that $w_j^i,w_j^p,w_j^{q}$ are on the segment of the rightward, leftward, and upward edge, respectively,
and are of equidistance~\settc{$10$} to each other.
See Figure~\ref{fig:set-gad} for an illustration.

\subparagraph{The edge- and the enforcement gadget.}\label{sub:enforcement}
For each edge~$e=\{u_i,w_j\}$ in~$G(I)$, we create $\enn$ (a constant value to be determined later) copies of the triple~\myemph{$A^j_i[z]=\{\alpha^j_i[z], \beta^j_i[z], \gamma^j_i[z]\}$}, $1\leq z \leq \enn$, of agents and embed them around the line segments of edge~$e$ in the grid (refer to \Cref{fig:edge-gadget}).
To connect to the set-gadget, we merge agent~$\gamma^j_i[{\enn}]$ and set-agent~$w^i_j$ together.
For technical reasons, we also use \myemph{$\gamma^j_i[0]$} to refer to~$u_i^j$.
To define the distances, let $\varepsilon_1,\varepsilon_2,\ldots, \varepsilon_{2\enn}$ be a sequence of increasing positive values with $2(2\enn-1)/(2\enn+1)\le \varepsilon_{2\enn-1} \le 2(2\enn-1)/(2\enn) < \varepsilon_{2\enn} = 2-\varepsilon$.
Now, we embed the newly added agents so that the distances between ``consecutive agents'' on the line increase with $z\in [\enn]$:
\begin{itemize}%
  \item The distance between agents~$\alpha^j_i[z]$ and $\beta^j_i[z]$ (marked in blue) is close to zero.
  \item The distance between agents~$\alpha^j_i[z]$  (resp.\ $\beta^j_i[z]$) and $\gamma^j_i[z]$ is $8+\varepsilon_{2z}$.
  \item The distance between $\alpha^j_i[{z}]$  (resp.\ $\beta^j_i[{z}]$) and $\gamma^j_i[z-1]$ is $8+\varepsilon_{2z-1}$.
\end{itemize}
In this manner, we will ensure that either all $A^j_i[z]$, $z\in [\enn-1]$,
or all $\{\gamma^j_i[z-1], \alpha^j_i[z], \beta^j_i[z]\}$, $i\in [\enn]$ belong to a \dsm[3] (to be proved later).

To determine the value~$\enn$, let the lengths of the segments for edge~$\{u_i,w_j\}$ in the orthogonal drawing of graph~$G(I)$ be $L_1$ and $L_2$, respectively; $L_2$ is zero if there is only one straight segment.
We set $\enn$ to the largest value satisfying $\sum_{z=1}^{2\enn}(8+0.01\cdot z) \le L_1+L_2$,
which is clearly a constant.
For brevity's sake, when using $\enn$, we mean the constant associated to an edge~$\{u_i,w_j\}$ in the drawing which will be clear from the context.
It is also fairly straightforward to check that one can choose the sequence~$\varepsilon_{i}$ so that the bending point of the chain is some agent~$\gamma^j_i[z']$, $z' \in [n-1]$ as shown in \Cref{fig:edge-gadget}.

By the construction of the gadgets above,
each set-agent~$w^i_j$ strictly prefers triple~$A^j_i[{\enn}]$ to triple~$\{w^i_j,w^p_j,w^q_j\}$ since $\dist(w^i_j, x) < \settc{10} = \dist(w^i_j, y)$ for all $x\in \{\alpha^j_i[\enn], \beta^j_i[\enn]\}$ and $y \in \{w^p_j, w^q_j\}$; recall that $w^i_j = \gamma^j_i[\enn]$ and $\dist(x, \gamma^j_i[\enn]) = 10 -\varepsilon$.
To ensure that exactly one of the two triples is chosen, we make use of the star-gadget from \cref{ex:star}.
More precisely, we introduce an agent triple \myemph{$H_j^i=\{f^i_j,g^i_j,h^i_j\}$} and embed them in such a way that the distances between two ``consecutive'' agents on the line towards the star-gadget increase:
\begin{itemize}%
  \item The distance between $f^i_j$ and $g^i_j$ is close to zero.
  \item The distance between agent~$h^i_j$ and each of $\{f^i_j,g^i_j\}$ is $10+2\varepsilon$.
  \item The distance between $f^i_j$ (resp.\ $g^i_j$) and each of~$A^j_i[{\enn}]$ is in range~$[10+\varepsilon, 10+2\varepsilon)$.
\end{itemize}
This means that the most preferred triple of agent~$h_j^i$ is $H_j^i$,
while both $f^i_j$ and $g^i_j$ prefer triple~$S$ to $H^i_j$ where $S = \{f_j^i,g_j^i,x\}$ and $x \in A^j_i[{\enn}]$.

Finally, we create $12$ agents, namely, $W=\{0,\ldots, 11\}$, according to \cref{ex:star} such that agents~$10$ and $11$'s most preferred triple is $\{10,11,h^i_j\}$, followed by $\{5,10,11\}$.
More precisely:
\begin{itemize}
\item The distance between agent~$10$ (resp.\ agent~$11$) and $h_j^i$ is $10+3\varepsilon$.
\item The distance between agent~$10$ (resp.\ agent~$11$) and $5$ is $10+4\varepsilon$.
\item The five agents from $\{0,\ldots,4\}$ form a regular pentagon with edge length~$a$.
Each two agents on the pentagon form with a private agent a triangle with edge lengths $a$ (marked in green), $b$ (marked in red), and $c$.
We set $b=10.1$ and $c=10.2$.
The length of the diagonal of the pentagon is~$\ell$.
\end{itemize}
Altogether, the lengths satisfy the relation $a<b<c<\ell$ and the specific angle~$\theta$ is at most 90 degrees.
Due to the chain, including $f^i_j$, $g^i_j$, and $h^i_j$, the distance from every agent not from~$W\cup \{h^i_j\}$ to every agent from~$W$ is larger than~$\ell$. %
We call the gadget, consisting of the star-agents and the triple~$H_j^i$,
the \myemph{star-gadget} for set-agent~$w_j^i$ and element-agent~$u_i^j$. 
\cref{fig:edge-gadget} provides an illustration of how the element-gadget, the set-gadget, and the star-gadget are embedded.
Note that since the angle between any two line segments is 90 degrees and the line segment has length at least $200$, we can make sure that such embedding is feasible.

This completes the description of the construction, which clearly can be done in polynomial time.
In total, we constructed $O(4\cdot 3n+3\cdot 3n+ 3\cdot 2\enn\cdot 3n + 15\cdot 3n) = O(n)$ agents. 
Note that we only need to have a good approximation of the embedding of the agents in the star-gadget and the equilateral triangle.

\subsection{The correctness proof for $\di=3$}

Before we proceed with the correctness proof,
we summarize the preferences derived from the embedding via the the following observation.
\begin{observation}[\appsymb]
  \label{obs:constr} 
  For each element~$i\in X$ and each set~$S_j\in \mathcal{S}$ with $S_j=\{i,p,q\}$, %
  let $0,\ldots,11$ denote the $12$ agents in the associated star-gadget.
  Then, the following holds.
\begin{enumerate}[(i)]
  \item\label{obs:11_12} The preference list of each agent~$x\in \{10, 11\}$ satisfies $\{h_j^i, 10, 11\} \succ_{x} \cdots$.
  \item\label{obs:11_12_hfg}

  For each triple~$B\neq \{h^i_j, f_j^i, g_j^i\}$ with $B  \succeq_{h^i_j}\{h_j^i, 10, 11\}$ it holds that
  $B \cap \{10,11\} \neq \emptyset$.

  \item\label{obs:fg} For each agent~$x\in \{f_j^i, g_j^i\}$ and each triple~$B\neq \{f_j^i,g_j^i,h_j^i\}$:
  \newH{\begin{itemize}
    \item[--] If $B=\{f_j^i,g_j^i,y\}$ (where $y\in \{\alpha_i^j[\enn], \beta_j^i[\enn], \gamma_j^i[\enn]\}$),
    then $B\succ_x \{f_j^i,g_j^i,h_j^i\}$.
    \item[--] If $B \succeq_x \{f_j^i,g_j^i,h_j^i\}$, then $B=\{f_j^i,g_j^i,y\}$ for some $y\in \{\alpha_i^j[\enn], \beta_j^i[\enn], \gamma_j^i[\enn]\}$.
  \end{itemize}}

  \item  \label{obs:Tz_c} For each~$z\in [\enn]$ the preference list of agent~$\gamma^j_i[z]$ satisfies $\{\alpha_i^j[z]$, $\beta_i^j[z]$, $\gamma_i^j[z]\} \succ_{\gamma_i^j[z]} \cdots$.
 \item \label{obs:Tz_ab}
 For each $z\in [\enn]$ the preference list of each agent~$x\in \{\alpha_i^j[z], \beta_i^j[z]\}$ satisfies 
 $\{\alpha_i^j[z]$, $\beta_i^j[z]$, $\gamma_i^j[z-1]\} \succ_x  \{\alpha_i^j[z], \beta_i^j[z], \gamma_i^j[z]\} \succ_x \cdots$.

\item \label{obs:Tz_c-a-b}

For each~$z\in [\enn-1]$ and each triple~$B\neq \{\alpha_i^j[z+1], \beta_i^j[z+1], \gamma_i^j[z]\}$ with 
$B \succeq_{\gamma_i^j[z]} \{\alpha_i^j[z+1], \beta_i^j[z+1], \gamma_i^j[z]\}$ it holds that $B \cap \{\alpha_i^j[z], \beta_i^j[z]\} \neq \emptyset$.

\item \label{obs:pref_w1}
For each~$B\neq \{w_j^i, w_j^p, w_j^q\}$ with $B \succeq_{w_j^i} \{w_j^i, w_j^p,w_j^q\}$ we have $B\cap \{\alpha_j^i[\enn], \beta_j^i[\enn]\} \neq \emptyset$.
\end{enumerate}
\end{observation}
\appendixproof{obs:constr}
{
  \begin{proof} 
  Statements~\eqref{obs:11_12} follows from the fact that the distance between agents~$10$ and $11$ is close to zero and agent~$h^i_j$ is the next nearest neighbor of them.
  Statement~\eqref{obs:11_12_hfg} follows from the fact that the closest neighbors of~$h^i_j$ are $f_j^i$ and $g_j^i$, followed by agents~$10$ and $11$.

  The first part of Statement~\eqref{obs:fg} is straightforward to verify after noticing that
  $\dist(x, y) < 10+2\varepsilon = \dist(x,h_j^i)$ for all $x\in \{f^i_j,g^i_j\}$ and $y \in \{\alpha^j_i[\enn], \beta_i^j[\enn], w^i_j\}$.  
  To prove the second part of Statement~\eqref{obs:fg}, assume that $B\succeq_x \{f^i_j,g^i_j,h^i_j\}$ with $B\neq \{f^i_j,g^i_j,h^i_j\}$.
  We first show that $\{f_j^i,g_j^i\}\subseteq B$.
  Notice that the distance between $f_j^i$ and $g_j^i$ is close to zero.
  Moreover, by construction, the next nearest neighbor of $f_j^i$ (resp.\ $g_j^i$) is at distance at least~$10+\varepsilon$.
  Since $\dist(f_j^i, \{f_j^i, g_j^i, h_j^i\}) = \dist(g_j^i, \{f_j^i, g_j^i, h_j^i\}) = 10+2\varepsilon + \dist(f_j^i, g_j^i) < 2(10+\varepsilon)$, any triple that is weakly preferred to $\{f_j^i, g_j^i, h_j^i\}$ by agent~$f_j^i$ (resp.\ $g_j^i$) must contain both $f_j^i$ and $g_j^i$.
  Now it is straight-forward to verify that $B\cap \{\alpha_i^j[\enn], \beta_i^j[\enn], \gamma_i^j[\enn]\}\neq \emptyset$ since $\alpha_i^j[\enn], \beta_i^j[\enn]$, and $\gamma_i^j[\enn]$ are the only agents to which $x$ has distance no more than $10+2\varepsilon$.

  Statement~\eqref{obs:Tz_c} follows directly from the fact that $\alpha_i^j[z]$ and $\beta_i^j[z]$ are the unique nearest neighbors of~$\gamma_i^j[z]$. %
  The reasoning for Statement~\eqref{obs:Tz_ab} is similar to the one for Statement~\eqref{obs:11_12}.
  The reasoning for Statement~\eqref{obs:Tz_c-a-b} is similar to the one for Statement~\eqref{obs:11_12_hfg}.
  
  To prove Statement~\eqref{obs:pref_w1}, we only need to observe that all agents in~$\{\alpha^j_i[\enn], \beta^j_i[\enn]\}$ are the nearest neighbors of~$w_j^i$, followed by the agents in~$\{w^p_j,w^q_j\}$.
\end{proof} 
}
\noindent
Finally, we show the correctness, i.e., ``$I=(X,\mathcal{S})$ admits an exact cover
if and only if the constructed instance admits a \dsm[3]''
via the following lemmas. \cref{lem:d=3-X3C->Stable} shows the ``only if''  direction and \cref{lem:d=3-Stable->X3C} the other. %

\begin{lemma}[\appsymb]\label{lem:d=3-X3C->Stable}
  If $\mathcal{K}\subset \mathcal{S}$ is an exact cover of $I$,
  then the following \dm[3]~$\Pi$ is stable.
\begin{itemize} %
  \item For each $S_j\in \mathcal{K}$ with $S_j=\{i,p,q\}$ add $\{w_j^i,w_j^p,w^q_j\}$ to $\Pi$.

  \item For each element~$i\in X$ and each set~$S_j\in\mathcal{S}$ with $i\in S_j$, call the agents in the associated star-gadget along with the tail agents $0,\ldots, 11, h_j^i, f_j^i$, and $g_j^i$.
  \begin{itemize}
    \item Add $H^i_j$, $\{5,10,11\}$, $\{1,6,8\}$, $\{2,3,7\}$, and $\{0,4,9\}$ to~$\Pi$.
    \item If $S_j\in\mathcal{K}$, then add all triples $\{\alpha^j_{i}[z], \beta^j_i[z], \gamma^j_i[z-1]\}$, $z\in [\enn]$, to $\Pi$.
    Otherwise, add all triples $A^j_i[z]$, $z\in [\enn]$, to $\Pi$.
\end{itemize}
\item For each element $i\in X$ let $S_{k}, S_{\setind}$ be the two sets which contain~$i$, but are not chosen in the exact cover~$\mathcal{K}$. Add $\{u_i,u_i^k,u_i^{\setind}\}$ to~$\Pi$.  
\end{itemize}
\end{lemma}

\appendixproof{lem:d=3-X3C->Stable}{
  \begin{proof}
    We prove this by showing that no agent can be involved in a blocking triple.

First, observe that if $S_j \in \mathcal{K}$ with $S_j=\{i,p,q\}$,
then $T=\{w_j^i,w_j^p,w^q_j\}\in \Pi$ and hence no agent in~$T$ is involved in a blocking triple since 
for each~$z\in S_j$ and each $B$ with $B\succ_{w^z_j} T$, it holds by Observation~\ref{obs:constr}\eqref{obs:pref_w1},
that $B\cap \{\alpha^j_z[\enn], \beta^j_z[\enn]\}\neq \emptyset$,
but both $\alpha^j_z[\enn]$ and $\beta^j_z[\enn]$ have been assigned to their most preferred triple, namely $\{\alpha^j_z[\enn], \beta^j_z[\enn], \gamma^j_z[\enn-1]\}$ (see Observation~\ref{obs:constr}\eqref{obs:Tz_ab}) as defined above.
If $S_j \notin \mathcal{K}$, to prove that $w_j^i$ is not involved in a blocking triple, we later show that $\gamma^i_j[\enn]$ (which is $w_j^i$) is not involved in a blocking triple.

We show that a blocking triple cannot contain any agent from $A_i^j[z]$ for all~$z \in [\enn]$,
where $\enn$ is a constant as defined in the construction.
We distinguish between two cases based on whether $S_j$ is in $\mathcal{K}$; again let $S_j=\{i,p,q\}$. %
\begin{description}
  \item[\textbf{Case 1: $S_j \in \mathcal{K}$}.] Then, for each $z\in [\enn]$,
  none of $\alpha_i^j[z]$ and $\beta_i^j[z]$ is involved in a blocking triple since both are matched to their most preferred triple (see Observation~\ref{obs:constr}\eqref{obs:Tz_ab}).
  Therefore, $A_i^j[z]$ is not blocking.
  For all $z \in [\enn-1]$, agent~$\gamma_i^j[z]$ is not in any blocking triple
  since every triple that is strictly preferred to~$\Pi(\gamma_i^j[z])$ by $\gamma_i^j[z]$ also contains $\alpha_i^j[z]$ or $\beta_i^j[z]$ (see \cref{obs:constr}\eqref{obs:Tz_c-a-b}) but none of the latter two is involved in a blocking triple; recall that we already showed that $\gamma^j_i[\enn]=w^i_j$ is not involved in any blocking triple.

\item[\textbf{Case 2: $S_j \notin \mathcal{K}$}.] Then, for each $z \in [\enn]$, agent~$\gamma_j^i[z]$ is matched to its most preferred triple $A_i^j[z]$ and hence, not involved in a blocking triple.
To show that neither $\alpha_i^j[z]$ nor $\beta_i^j[z]$ is involved in a blocking triple for each $z\in[\enn]$, let us consider an arbitrary agent~$x\in \{\alpha_i^j[z], \beta_i^j[z]\}$, $z\in [\enn]$.
By \cref{obs:constr}\eqref{obs:Tz_ab}, triple~$\{\alpha^j_i[z], \beta^j_i[z], \gamma^j_i[z-1]\}$ is the only triple which is preferred to $\Pi(x)=A^j_i[z]$ by agent~$x$.
However, $\gamma^i_j[z]$ does not prefer $\{\alpha^j_i[z], \beta^j_i[z], \gamma^j_i[z-1]\}$ to her assigned triple since she is already matched to her most preferred triple %
(recall that $\gamma^j_i[0]=u^j_i$ and by construction, agent~$\Pi(u_i^j)$ is also the most preferred triple of~$u^j_i$).
This means that $x$ cannot be involved in a blocking triple.
 \end{description}
 
\noindent Further, for each element $i\in X$ with $i\in S_{k}, S_{\setind}$ and $S_{k}, S_{\setind} \notin \mathcal{K}$, the agents $u_i^k,u_i^{\setind}$ are matched to their most preferred coalition and so is $u_i$ and they do not form blocking.

 It remains to show that no agent from $H_j^i\cup \{0,\ldots,11\}$ is involved in a blocking triple.
 Notice that agent~$h_j^i$ is matched to her most preferred triple $H_j^i$ and hence, not involved in a blocking triple.
 Neither $f_j^i$ nor $g_j^i$ is involved in a blocking triple because of the following:
 All triples that are weakly preferred to $\Pi(f_j^i)=\Pi(g_j^i)=H_j^i$ by $f_j^i$ or $g_j^i$ contain $f^i_j$ and $g^i_j$, %
 and someone from $A^j_i[\enn]$ (see the second part of \cref{obs:constr}\eqref{obs:fg}),
 but we just showed that no agent from $A^j_i[\enn]$ is involved in a blocking triple.
 
 From the construction and by Observation~\ref{obs:constr}\eqref{obs:11_12}, both $10$'s and $11$'s most preferred triple is $\{10, 11, h_j^i\}$ followed by $\{5,10,11\} = \Pi(10)=\Pi(11)$.
 Since $h_j^i$ is not in a blocking triple, $\{10, 11, h_j^i\}$ is not blocking.
 Thus, $10$ and $11$ are not involved in a blocking triple.

 Finally, by construction, agents~$5$,~$7$, and $9$ are not involved in a blocking triple since they are each matched to their most preferred triple, respectively.
 Neither is agent~$2$ involved in a blocking triple since she only prefers~$\{1,2,3\}$ to her assigned triple~$\{2,3,7\}$ but agent~$3$ does not prefer $\{1,2,3\}$ to her assigned triple~$\{2,3,7\}$.
 Agent~$3$ cannot be involved in a blocking triple since the triple which she prefers to her assigned one involves either agent~$2$ or agent~$4$ but we just reasoned that neither~$2$ nor~$4$ is involved in a blocking triple. 
 Similarly, we can infer that agent~$4$ is not involved in blocking triple:
 The triple which agent~$4$ prefers to her assigned one involves either $4$ or $5$ but we just reasoned that neither~$4$ nor~$5$ is involved in a blocking triple. 
 Finally, none of $\{1,6,8\}$ is involved in a blocking triple since no other agent from the same star-gadget or from the element-gadget, set-gadget, or edge-gadget is involved in blocking triple and all agents not from the same star-gadget are further away.
 Summarizing, $\Pi$ is stable.
\end{proof}
}
The proof of the other direction is based on the following properties.
\begin{lemma}[\appsymb]\label{lem:set-agent}
 Let $\Pi$ be a \dsm[3] of the constructed instance.
  For each element~$i\in X$ and each set~$S_j$ with $S_j=\{i,p,q\}$, the following holds:
 { \begin{enumerate}[(i)]
   \item\label{lem:fgh} $H_j^i\in \Pi$. %
  \item\label{lem:left-right} $\Pi$ contains either all triples~$\{\alpha^j_i[z],\beta^j_i[z],\gamma^j_i[z]\}$ or all triples~$\{\alpha^j_i[z], \beta^j_i[z],\gamma^j_i[z-1]\}$,  $z\in [\enn]$.
\end{enumerate}
}
\end{lemma}

\appendixproof{lem:set-agent}{
\begin{proof}
  In this proof, since $i$ and $j$ are fixed, we drop $i$ and $j$ from the superscripts and subscripts (except for~$w^j_i$) for ease of notation.
  Further, let $0,\ldots,11$ denote the agents of the star-gadget associated to the set-agent~$w_j^i$.
  We first prove the following two claims which will be used in the proof. 
\begin{claim}\label{claim:51011}
  $\Pi$ contains $\{5,10,11\}$. %
\end{claim}

\appendixproof{claim:51011}
{
  \begin{proof}[Proof of \cref{claim:51011}]
  \renewcommand{\qedsymbol}{$\diamond$}
   This follows from the inclusion follows from \cref{lem:pentagon} and from the fact that the distance of each agent not from~$\{0,\ldots, 11\}$ to every agent in~$\{0,\ldots,9\}$ is larger than the length of the diagonal of the regular pentagon. 

 \end{proof}
}

   Now, to show the first statement, suppose, for the sake of contradiction, that %
  $\Pi(h)\neq \{f,g,h\}$.
  Recall that $\{5,10,11\} \in \Pi$ (see Claim~\ref{claim:51011}).
Since $\{h,10,11\}$ is the most preferred triple for both agents~$10$ and $11$ (see \cref{obs:constr}\eqref{obs:11_12}), by stability, $\Pi(h) \succeq_h \{h,10,11\} $.
Then, by Observation~\ref{obs:constr}\eqref{obs:11_12_hfg}, $\Pi(h) \cap \{10,11\} \neq \emptyset$, a contradiction to $\Pi$ being a partition.
This completes the proof.

The next claim states that there are exactly two possible matchings of the agents on the chain connecting the element- and set-gadget.
\begin{claim}\label{clm:zz-1}
  For each $z \in [\enn]$, if $A[z]\notin \Pi$, then $\{\alpha[z],\beta[z],\gamma[z-1]\}\in \Pi$.
\end{claim}

  \begin{proof}[Proof of \cref{clm:zz-1}]
  \renewcommand{\qedsymbol}{$\diamond$}
  Consider an arbitrary index~$z \in \newH{[\enn]}$ and assume that $A[z]\notin \Pi$.
  Notice that, by Observation~\ref{obs:constr}\eqref{obs:Tz_c}, triple~$A[z]$ is the unique most preferred triple of $\gamma[z]$; recall that $\gamma[\enn]=w_j^i$.
  Hence, by our assumption, to prevent $A[z]$ from blocking~$\Pi$,
  agent~$\alpha[z]$ or agent~$\beta[z]$ has to be matched in a triple that she weakly prefers to~$A[z]$.
  By Observation~\ref{obs:constr}\eqref{obs:Tz_ab}, triple $\{\alpha[z],\beta[z],\gamma[z-1]\}$
  is the only triple which is weakly preferred to~$A[z]$ by $\alpha[z]$  or $\beta[z]$.
  Therefore, $\{\alpha[z],\beta[z],\gamma[z-1]\} \in \Pi$.
\end{proof}

To show the lemma, we distinguish between two cases: %
  \begin{description}
    \item[{Case 1:} 
    {$\{\alpha[\enn], \beta[\enn], \gamma[\enn-1]\}\in \Pi$}.]
    This means that $A[\enn-1]\notin \Pi$.
    By \cref{clm:zz-1}, it follows that $\{\alpha[\enn-1], \beta[\enn-1], \gamma[\enn-2]\}\in \Pi$.
    By repeatedly using the above reasoning, we infer that $\{\alpha[z], \beta[z], \gamma[z-1]\}\in \Pi$ for all $z\in [\enn]$, as desired.

    \item[%
    {Case 2:} {$\{\alpha[\enn], \beta[\enn], \gamma[\enn-1]\} \notin \Pi$}.]
    Notice that $S\coloneqq \{\alpha[\enn], \beta[\enn], \gamma[\enn-1]\}$ is the unique most preferred triple of both $\alpha[\enn]$ and $\beta[\enn]$ (see \cref{obs:constr}\eqref{obs:Tz_ab}).
    By stability, we have that $\Pi(\gamma[\enn-1])\succeq_{\gamma[\enn-1]} S$.
    By \cref{obs:constr}\eqref{obs:Tz_c-a-b}, we infer that $\Pi(\gamma[\enn-1])\cap \{\alpha[\enn-1], \beta[\enn-1]\} \neq \emptyset$.
    This implies that $\{\alpha[\enn-1], \beta[\enn-1], \gamma[\enn-2]\}\notin \Pi$,
    and by the contra-positive of \cref{clm:zz-1} we have that $A[\enn-1]\in \Pi$.
    By repeatedly using the above reasoning, we infer that $A[z]\in \Pi$ for all $z\in [\enn-1]$.
    It remains to show that $A[\enn]\in \Pi$.
    This is straightforward to see since $A[\enn-1]\in \Pi$, meaning that $\gamma[\enn-1]$ is not available anymore:
    By \cref{obs:constr}\eqref{obs:Tz_c}--\eqref{obs:Tz_ab}, we infer that $A[\enn]\in \Pi$ is the unique most preferred triple of each agent in~$A[\enn]$.
    Hence, $A[\enn]\in \Pi$.
  \end{description}
  This completes the proof of \cref{lem:set-agent}.
\end{proof}
}

Now, we consider the ``if'' direction.
\begin{lemma}\label{lem:d=3-Stable->X3C}
  If $\Pi$ is a \dsm[3], then the subcollection~$\mathcal{K}$ with $\mathcal{K}=\{S_j \in \mathcal{S}\mid
\{\alpha^j_i[1],\beta^j_i[1],\gamma^j_i[0]\} \in \Pi \text{ for some } i \in S_j\}$ is an exact cover.
\end{lemma}

  \begin{proof}
First of all, for each two chosen~$S_j, S_k\in \mathcal{K}$ we observe that it cannot happen that
$S_j\cap S_k\neq \emptyset$ as otherwise $\{u_i,u_i^j,u_i^k\}$ is a blocking triple; recall that $\gamma^j_i[0]=u_i^j$ and $\gamma^k_i[0]=u_i^k$.
It remains to show that $\mathcal{K}$ covers each element at least once.

Now, for each element~$i\in X$, let $S_j,S_k,S_{\setind}$ denote the three sets that contain~$i$.
We claim that at least one of $S_j,S_k,S_{\setind}$ belongs to~$\mathcal{K}$ because of the following.
If $S_j\notin \mathcal{K}$, then by construction, it follows that $T=\{\alpha^j_i[1],\beta^j_i[1],\gamma^j_i[0]\}\notin \Pi$.
By \newH{\cref{lem:set-agent}\eqref{lem:left-right}}, it follows that $A^j_i[1]\in \Pi$.
Since $T$ is the most-preferred triple of both $\alpha^j_i[1]$ and $\beta^j_i[1]$ (see \cref{obs:constr}\eqref{obs:Tz_ab}),
by stability, $u_i^j$ must be matched in a triple which she weakly prefers to~$T$. %
Since $A^j_i[1]\in \Pi$, it follows that either $\{u^j_i,u_i^{k}, u^{\setind}_i\} \in \Pi$ or $\{u^j_i,u_i,v\}\in \Pi$ for some $v\in \{u_i^k,u^{\setind}_i\}$.
It cannot happen that  $\{u^j_i,u_i^{k}, u^{\setind}_i\} \in \Pi$ as otherwise there will be at least three blocking triples, including $\{u_i,u_i^j,u^{\setind}_i\}$.
Hence, $\{u^j_i,u_i,v\}\in \Pi$ for some $v\in \{u_i^k,u^{\setind}_i\}$.
Without loss of generality, assume that $v=u^k_i$.
Then, it is straightforward to check that $\{u^{\setind}_i, \alpha^{\setind}_i[1], \beta^{\setind}_i[1]\}\in \Pi$.
This implies that $S_{\setind}\in \mathcal{K}$.

{To complete the correctness proof, we show that for each element~$p\in S_{\setind}\setminus \{i\}$ it holds that $\{\alpha_{p}^r[1], \beta_{p}^r[1], \gamma_p^r[0]\}\in \Pi$.
  Let $S_{\setind}=\{i,p,q\}$.
  Since $S_{\setind}\in \mathcal{K}$, by definition and by \cref{lem:set-agent}\eqref{lem:left-right},
  we infer that $\{\alpha_i^{\setind}[\enn], \beta_i^{\setind}[\enn], \gamma_{i}^{\setind}[\enn-1]\}\in \Pi$ (for some constant $\enn$ defined in the construction).
  We infer that $\{w_{\setind}^i,w_{\setind}^p,w_{\setind}^q\} \in \Pi$ due to the following: By \cref{lem:set-agent}\eqref{lem:fgh}, we know that $H_{\setind}^{i}\in \Pi$; recall that $H_{\setind}^i=\{f_{\setind}^i, g_{\setind}^i,h_{\setind}^i\}$.
  Since both $f_{\setind}^i$ and $g_{\setind}^i$ prefer $\{f_{\setind}^i, g_{\setind}^i,w_{\setind}^i\}$ to $H_{\setind}^i$ (see the first part of \cref{obs:constr}\eqref{obs:fg}),
  it follows by stability that $\Pi(w_{\setind}^i)\succeq_{w_{\setind}^i} \{f_{\setind}^i, g_{\setind}^i,w_{\setind}^i\}$.
  By \cref{obs:constr}\eqref{obs:pref_w1}, we infer that $\Pi(w_{\setind}^i) = \{w_{\setind}^i,w_{\setind}^p,w_{\setind}^q\}$ since $\alpha_i^{\setind}[\enn]$ and $\beta_{i}^{\setind}[\enn]$ are not available anymore.
  This means that $A_{p}^{\setind}[\enn'], A_{q}^{\setind}[\enn'']\notin \Pi$ since $w_{\setind}^p = \gamma^{\setind}_p[\enn']$ and $w_{\setind}^q = \gamma^{\setind}_p[\enn'']$ (for some constants~$\enn'$ and $\enn''$).
  Consequently, we infer by \cref{lem:set-agent}\eqref{lem:left-right} that $\{\alpha_p^{\setind}[1], \beta_p^{\setind}[1], \gamma_p^{\setind}[0]\}, \{\alpha_q^{\setind}[1], \beta_q^{\setind}[1], \gamma_q^{\setind}[0]\} \in \Pi$, as desired.
}
\end{proof}
This concludes the proof of \cref{thm:main} for $\di=3$.

\section{\edsr{} with $\di \ge 4$}\label{sec:multi}
In this section we look at the cases where $\di \ge 4$, and let $\hs\coloneqq \lfloor (\di-1)/2 \rfloor$. %
The general idea of the reduction is similar to the case where $\di =3$, and we still reduce from \xcts{}.
Briefly put, we adapt the star-gadget from \cref{ex:star}.
However, depending on whether $\di$ is even or not, we need to carefully revise the star-gadget from \cref{ex:star} to make sure the enforcement gadget works.
We will replace each pentagon-agent with a subset of agents of size~$\hs$, and each further agent from the triangle with two agents if $\di$ is even.
We also need to update both the replacement and the enforcement gadget.
In \cref{sub:construction-G}, we describe in detail what the new star-gadgets and the the remaining gadgets look like, and how they are connected to each other.
In \cref{sub:correctness-G} we show the correctness.

\appendixsection{sub:construction-G}
\subsection{The construction}\label{sub:construction-G}
 
   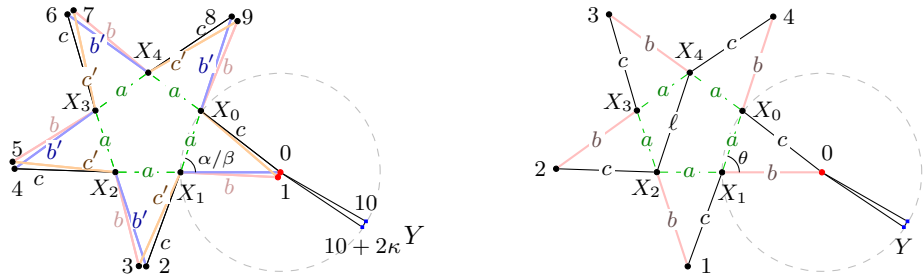
\begin{figure}
     \centering
     \begin{subfigure}[t]{0.45\linewidth}
    \centering
  \begin{tikzpicture}[scale=.13]

  \begin{scope}%
    \dpentevenN{6.6}{10.1}{10.2}{0}{10.1}{10.2}%

    \dpentevennames{6.6}{0}

    \dnamesneven{6.6}{0}

    \drawab
  \end{scope}
  \begin{scope}
    \node[dnode] at ($(n1)+(-30:10)$) (x) {};
    \node[dnode] at ($(n1)+(-34:10)$) (y) {};
    \node[above=0pt of x] {\footnotesize $10$};
    \node[below=0pt of y] {\footnotesize $10+2\hs$};
    \node at ($(n1)+(-25:15)$) {$Y$};
    \draw (n1) -- (x);
    \draw (n1) -- (y);
    \draw (x) -- (y);
  \end{scope}
\end{tikzpicture}
\end{subfigure}~~~\begin{subfigure}[t]{0.45\linewidth}
    \centering
  \begin{tikzpicture}[scale=.13]

  \begin{scope}%
    \dpentN{6.6}{10.1}{10.2}{0}{10.1}{10.2}
    \dpentnames{6.6}{0}

    \dnamesn{6.6}{0}

    \drawphi
  \end{scope}
  \begin{scope}
    \node[dnode] at ($(n1)+(-30:10)$) (x) {};
    \node[dnode] at ($(n1)+(-34:10)$) (y) {};
    \node[above=0pt of x] {};
    \node[below=0pt of y] {\footnotesize $Y$};
    \draw (n1) -- (x);
    \draw (n1) -- (y);
    \draw (x) -- (y);
  \end{scope}
\end{tikzpicture}
\end{subfigure}
\caption{A star-structured instance adapted from Arkin et al.~\cite{arkin2009geometric}, similar to \cref{ex:star}. The left one is for even~$\di$, while the right one is for odd~$\di$, both described in \cref{ex:heptagon_n-DSM}. See the caption of \cref{ex:star} for further explanation regarding the colors of the edges.} %
\label{fig:pentagon_n-DSM}
\end{figure}
We first describe the adapted star-gadgets through the following example (also see \cref{fig:pentagon_n-DSM}).

\begin{example}\label{ex:heptagon_n-DSM}
  We first consider the construction for even~$\di$, i.e., $\di=2\hs+2$.
  Consider an instance with $7\hs+11$ agents called $W$ where $5\hs$ agents are embedded as the five vertices of a pentagon with $\hs$ agents at each vertex of the pentagon.
We denote the five sets of points at the five vertices of the pentagon as $X_0, \ldots, X_4$.
All points in each cluster~$X_i$, $0\le i \le 4$, are embedded within an enclosing circle of radius close to zero, \newH{with the intention that 
  a \dm{} is stable only if all agents in~$X_i$ are matched together.} 
For each $i \in \{ 0, \ldots, 4\}$, the distance between each point~$X_i$ and each point in~$X_{i+1 \bmod 5}$ is in the range of $[a, a+\varepsilon_\di]$,
while the distance between each point in~$X_i$ and each point in~$X_{i+2 \bmod 5}$ is
in the range of $[\ell, \ell+\varepsilon_\di]$. %
There are $10$ points $\{0,\ldots,9\}$ that form a star with the pentagon, as shown in \Cref{fig:pentagon_n-DSM}~(left).
For each $i\in \{0,\ldots,4\}$, embed the points~$2i$ and $2i+1$ as follows:
point~$2i$ is at a distance between $c$ and $c+\varepsilon_\di$ to every point in~$X_i$,
and at a distance between $b'$ and $b'+\varepsilon_\di$ to every point in~$X_{i+1 \bmod 5}$.
Point $2i+1$ is at a distance between $c'$ and $c'+\varepsilon_\di$ to every point in~$X_i$,
and at a distance between $b$ and $b+\varepsilon_\di$ to every point in~$X_{i+1 \bmod 5}$.
Finally, the distance~$\dist(2i, 2i +1)$ is close to $0$.
Here the mentioned values satisfy the following relations $a< b < c < \ell$, $b< b'<\ell$, $c<c'<\ell$, $b+ b' <3a$, $c + c'< 3a$, $b+b' < a+\ell$, and $c+c' < b+\ell$.

The remaining $2\hs+1$ points, denoted by $10, \ldots, 10+2\hs$ in the figure, are called $Y$; note that $|Y|=2\hs + 1 = \di-1$.
Together, $W \coloneqq \bigcup\limits_{i\in\{0,\ldots, 4\}} X_i \cup \{0,\ldots, 9\} \cup Y$.
All points in $Y$ are embedded within an enclosing ball with radius close to zero. %
  For each point~$y$ in~$Y$, it holds that~$b-\varepsilon_{\di} \le \dist(0,y) < b$
  and $b-\varepsilon_{\di}\le \dist(1,y) < b$, and %
for each each point~$w$ in $W \setminus (\{0,1\} \cup Y)$
it holds that $\dist(w,y) > \ell$. %
Points $0$ and $1$ are the two points from~$W\setminus Y$ which are closest to the points in~$Y$.
  To specify the embedding, We first fix points $0$, $1$, and $Y$ such that the distances between them are as stated above and they are embedded roughly around a straight line.
Then, we fix the positions of $X_0$, $X_1$, and the centroid of the pentagon to ensure the values~$a,b,b',c,c'$, and $\ell$ satisfy the above relations. %
For each $i\in \{0,1,2,3, 4\}$ and each two points~$x\in X_i$ and $x'\in X_{i+1\bmod 5}$, the angle $\alpha$ (resp.\ $\beta$) at the points $2i$, $x$, and $x'$ (resp.\ $2i+1$, $x'$, and $x$) is less than $90$ degrees.
The angle at points~$y$, $j$, and $x$ ($y\in Y$, $\{i,j\}=\{0,1\}$, $x\in X_i$) is more than $90$ degrees.
After fixing $X_0$, $X_1$, $0$, and $1$, we can determine the other points by simple calculations.

Now, we turn to odd $\di$, i.e., $\di=2\hs+1$. Instead of having ten points $\{0,\ldots,9\}$, we create five points that form a star with the pentagon. 
 Consider an instance with $7\hs + 5$ agents called $W$ where $5\hs$ agents are embedded to replace the five vertices of a pentagon with $\hs$ agents at each vertex of the pentagon.
  That is, each vertex of the pentagon is a cluster of points. note the five clusters of points by $X_0, X_1, X_2, X_3$, and $X_4$.
  There are five points $\{0,1,2,3,4\}$ that form a star with the pentagon, as in \Cref{ex:star} (see  \Cref{fig:pentagon_n-DSM}~(right)).
  Point $i$ is at a distance $b$ from $X_i$ and $c$ from $X_{i+1 \bmod 5}$, for each $i \in \{ 0, \ldots, 4\}$ where $a< b < c < \ell$ and $b<2a$. 

  The remaining $2\hs$ points are called $Y$. %
Together, $W \coloneqq \bigcup\limits_{i\in\{0,\ldots, 4\}} X_i \cup \{0,1,2,3, 4\} \cup Y$.
All points in $Y$ are embedded within an enclosing circle with radius close to zero. %
  For each point~$y$ in~$Y$, it holds that~$b-\varepsilon \le \dist(0,y) < b$, and %
for each each point~$w$ in $W \setminus (\{0\} \cup Y)$
it holds that $\dist(y,w) > \ell$. %
Point $0$ is the only point from~$W\setminus Y$ which is closest to the points in~$Y$.
The remaining unmentioned points are at distance at least $b/2$ to the points~$Y$.
 We specify the embeddings of the agents similarly to the one for even~$\di$.

Using a similar reasoning as to \cref{ex:star}, we claim that the above embeddings are feasible.

\end{example}

\noindent Since the distance between each two points in $X_i$ is close to zero, we assume it to be $0$ for ease of reasoning. %
 The following lemma summarizes the crucial effect of the star-gadget.

\begin{lemma}[\appsymb]\label{lem:heptagon4DSM}
  Every \dsm{} $\Pi$ of the instance in \Cref{ex:heptagon_n-DSM} satisfy
  that
 \newH{ if $\di$ is even, then $\Pi(0) \cap Y \neq \emptyset$ or $\Pi(1) \cap Y \neq \emptyset$,
  and if $\di$ is odd, then $\Pi(0) = Y\cup \{0\}$.}
\end{lemma}

\appendixproof{lem:heptagon4DSM}{\begin{proof}
  \noindent \textbf{The case when $\di$ is even.}
    Suppose, towards a contradiction, that $\Pi(0) \cap Y = \emptyset$ and $ \Pi(1) \cap Y = \emptyset$.
  To improve readability, in the following, the subscript~$i+1$ in $X_{i+1}$ is taken modulo $5$.
  Then, the most preferred available coalition for $0$ and $1$ is $X_0 \cup X_1 \cup \{0,1\}$. 
  Furthermore, for each~$i\in \{0,\ldots,4\}$,
  the distance between the points $2i$ and $2i +1$, $i \in \{1, \ldots, 4\}$, is close to zero,
  and the next closest points to $2i$ (resp.\ $2i+1$) are those from~$X_{i+1}$, followed by those from~$X_i$.
  Hence, for each $i \in \{0, \ldots, 4\}$,
  the most preferred available coalition of $2i$ (resp.\ $2i+1$) is $\{2i, 2i +1\} \cup X_{i+1} \cup X_i$.
  We will use this observation in the proof whenever we consider potential blocking coalitions containing $2i$ and $2i +1$. %

  First, we show that there exists an index~$i \in \{0, \ldots, 4\}$ such that $X_i \cup X_{i +} \subseteq \Pi(x)$ for some $x \in X_i$. That is at least one ``pentagon edge'' is matched together.

 We begin with the following simple claim.

\begin{claim}\label{obs:propPentagon4D}
  For each $i \in \{0, \ldots, 4\}$ 
  it holds that if $X_i \nsubseteq \Pi(x_i)$ for some $x_i\in X_i$, then $\distP{x_i} \geq (\hs+3)a$ for each~$x_i\in X_i$.
\end{claim}
\begin{proof}[Proof of \cref{obs:propPentagon4D}]
  \renewcommand{\qedsymbol}{$\diamond$}
  If $X_i \nsubseteq \Pi(x_i)$, then $|\Pi(x_i) \cap X_i| \leq \hs-1$ since $|X_i| = \hs$.
  Then at least $\hs+3$ points in $\Pi(x_i)$ are not in $X_i$. Hence, $\distP{x_i} \geq (\hs+3)a$.
\end{proof}

Whenever $X_i \subseteq \Pi(x_i)$ for some $x_i \in X_i$, we abuse the notation to write the matching coalition containing $X_i$ as $\Pi(X_i)$ for $i \in \{0, \ldots, 4\}$. 

Towards a contradiction, suppose that for each $i\in \{0,\ldots, 4\}$,
there exists a point~$x\in X_i$ such that that~$X_i \cup X_{i +1} \nsubseteq \Pi(x)$.
Without loss of generality, we consider~$i=3$ and two cases based on how $X_3$ is matched. We will show contradiction in both cases, thereby implying $X_i \cup X_{i +1} \subseteq \Pi(x)$ for some $x \in X_i$ for some $i \in \{0, \ldots, 4\}$.

\begin{itemize}
  \item[\textbf{Case 1:}] $X_3 \nsubseteq \Pi(x_3)$ for some $x_3 \in X_3$. Then from \Cref{obs:propPentagon4D}%
  , $\distP{x_3} \geq (\hs+3)a$.
Then, we observe the following.
\begin{enumerate}[(i)]
  \item In order for $X_2 \cup X_3 \cup \{4,5\}$ to not be blocking,
  $\distP{x_2} \leq \hs a + c+c'$.
  Recall that $c+c'< 3a$.
  By the contra-positive of \Cref{obs:propPentagon4D}, we have that $X_2 \subseteq \Pi(x_2)$ for some $x_2 \in X_2$. 

  \item In order for $X_3 \cup X_4 \cup \{6,7\}$ to not be blocking,
  \begin{align}
    \distP{x_4} \leq \hs a+ b+b'.\label{eq:dist-X4}
  \end{align}
  Recall that $b+b'< 3a$.
  By the contra-positive of \Cref{obs:propPentagon4D}, $X_4 \subseteq \Pi(x_4)$ for some $x_4 \in X_4$. 
\end{enumerate}
 
Since $\distP{x_4} \leq \hs a+ b+b'$ (see \eqref{eq:dist-X4}), and since by assumption not all points in $X_3$ can be matched together with $X_4$,
it follows that $X_0 \cap \Pi(X_4)  \neq \emptyset$. 

Recall that by our assumption, for all~$i\in \{0,\ldots,4\}$, it holds that $X_i \cup X_{i +1} \nsubseteq \Pi(x)$ holds for some $x \in X_i$.
Therefore, $X_0 \nsubseteq \Pi(X_4)$. Then, 
\[ \text{for each } x_0 \in X_0 \cap \Pi(X_4),\text{ it holds that } \distP{x_0} \geq \hs a +b+\ell \text{ and }\]
\[\text{ for each } x_0 \notin X_0 \cap \Pi(X_4), \text{ it holds that }\distP{x_0} \geq (\hs+3)a.\]

This implies that for all~$x_0\in X_0$, we have that $\distP{x_0} > \hs a +c+c'$.
Hence, for $X_0\cup X_1 \cup \{0,1\}$ to not be blocking, there must exist a point $x_1 \in X_1$ with $\distP{x_1} \leq \hs a + b+b' < (\hs+3)a$; recall that $\max(b+b', c+c') < 3a$.
Let $x_1\in X_1$ denote such a point, i.e., $\distP{x_1} \le \hs a + b + b'$.
Then, by the contra-positive of \Cref{obs:propPentagon4D}, we have that $X_1 \subseteq \Pi(x_1)$.
Since not all points in $X_0$ can be matched with $X_1$ and $\distP{x_1} \leq \hs a + b+b'$, we have that $X_2 \cap \Pi(X_1) \neq \emptyset$.
Recall that $X_2 \subseteq \Pi(X_2)$.
This implies that $X_1$ and $X_2$ are matched together, a contradiction.

\item[\textbf{Case 2:}] $X_3 \subseteq \Pi(x_3)$ for some $x_3 \in X_3$.
We distinguish two cases based on how $X_4$ is matched.

\begin{itemize}
\item[\textbf{Case 2.1:}]  $X_4\subseteq \Pi(x_4)$ for some~$x_4\in X_4$,
then no point in~$X_4$ is matched with $X_3$
and at most $\hs-1$ points from $X_0$ is matched with $X_4$.
This implies that
$\distP{x_4}\ge (\hs-1)a + b + b' + c > \dist(x_3, \{6,7\}\cup X_3\cup X_4)$ for all $x_4\in X_4$. 
Furthermore, at most $\hs-1$ points from $X_2$ can be matched with $X_3$,
implying that
$\distP{x_3}\ge (\hs-1)a + b + b' + c > \dist(x_3, \{6,7\}\cup X_4\cup X_3)$ for all $x_3\in X_3$.
This results in $X_3\cup X_4\cup \{6,7\}$ blocking~$\Pi$.

\item[\textbf{Case 2.2:}] If $X_4\nsubseteq \Pi(x_4)$ for some $x_4\in X_4$, then $\distP{x_4} \geq (\hs +3)a + c + c'$ by \Cref{obs:propPentagon4D}. This case is symmetric to Case 1 by replacing $X_3$ with $X_4$ as shown next.

In order for $X_0 \cup X_4 \cup \{8,9\}$ to not blocking~$\Pi$, there must exist an~$x_0 \in X_0$ with $\distP{x_0} \leq \hs a + b+b' < (\hs +3)a$ since $b+b' < 3a$.
By \cref{obs:proppentagon_nD}, we have that $X_0 \subseteq \Pi(x_0)$.
By assumption that no pentagon edge is matched with each other, we further infer that $\Pi(X_0) \cap X_1 \neq \emptyset$.

But since no ``pentagon edge'' is matched, it follows that $X_1 \nsubseteq \Pi(X_0)$. Therefore, by \Cref{obs:propPentagon4D}, we have that $\distP{x_1} \geq (\hs+3)a$ for each~$x_1 \in X_1$. Then, in order for coalition~$X_1 \cup X_2 \cup \{2,3\}$ to not be blocking,
there must exist a point~$x_2 \in X_2$ with $\distP{x_2} \leq \dist(x_2, X_1 \cup X_2 \cup \{2,3\}) = \hs a+ b+ b' < ( \hs +3)a$. 
Therefore, by \cref{obs:proppentagon_nD}, it follows that $X_2 \subseteq \Pi(x_2)$ for each~$x_2 \in X_2$ and $\Pi(X_2) \cap X_3 \neq \emptyset$ since $X_2$ cannot be matched with all~$\hs$ points in $X_1$.
Recall that we we assumed in this case that $X_3 \subseteq \Pi(x_3)$ for some $x_3 \in X_3$. Then  $X_2 \cup X_3$ are matched together, contradicting our assumption that no ``pentagon edge'' is matched together.

\end{itemize} 
\end{itemize}

Hence, we proved that $X_i \cup X_{i +1} $ are matched together for some $i \in \{0, \ldots, 4\}$.
Without loss of generality,  let $X_3 \cup  X_4 \cup \{u, v\}\in \Pi$. 
As in the proof of \Cref{lem:pentagon}, for each possible value of~$u$ and $v$ we will show a contradiction to stability of~$\Pi$,
by showing that no agents~$u$ and $v$ exist which are matched with $X_3 \cup X_4$ in a \dsm. 
We distinguish between several cases.
\begin{itemize}
\item[\textbf{Case 1:}] $\{u,v\} \nsubseteq X_0 \cup X_2 \cup \{ 4,5,6,7,8,9\}$.
Then, $ X_3 \cup X_4 \cup \{6,7\}$ is blocking. 

\item[\textbf{Case 2:}] $u \notin X_0 \cup X_2 \cup \{ 4,5,6,7,8,9\}$ and $v \in X_0 \cup X_2 \cup \{ 4,5,6,7,8,9\}$.
\begin{itemize}
  \item[\textbf{Case 2.1:}] $v \in X_0$.
  For each $x_3\in X_3$,
  it holds that $\distP{x_3} \ge \hs a + \ell + \ell > \hs a + c + c'$.   
 The first inequality follows because $u \notin X_0\cup X_2\{4,5,6,7,8,9\}$.
 The second inequality follows from the fact that $\ell >\max(c,c')$.
Again, since $u \notin X_0\cup X_2\cup \{4,5,6,7,8,9\}$, for each $x_4\in X_4$ it holds that
 $\distP{x_4}\ge \hs a + a + \ell > \hs + b + b'$; recall that $a+ \ell > b+b'$.
 Hence, $X_3\cup X_4\cup \{6,7\}$ is blocking.
 
\item[\textbf{Case 2.2:}]$v \in X_2$. %
Recall that $u\notin  X_0 \cup X_2 \cup \{ 4,5,6,7,8,9\}$.
For each $x_3 \in X_3$, it holds that $\distP{x_3} \geq \hs a +a+\ell >\hs a + b+b'$.
 In order for $X_2\cup X_3\cup \{4,5\}$ to not block,
   by the contra-positive of \Cref{obs:propPentagon4D},
   we have that
   $X_2\subseteq \Pi(x_2)$ for all $x_2\in X_2$.
   This is a contradiction to $v\in X_2$ and $u\notin X_2$.

\item[\textbf{Case 2.3:}] $v \in \{4,5,6,7,8,9\}$, then $ X_3 \cup X_4 \cup \{6,7\}$ is blocking since $u \notin \{6,7\}$. 
\end{itemize}

\item[\textbf{Case 3:}] $u \in X_0 \cup X_2 \cup \{ 4,5,6,7,8,9\}$ and $v \notin X_0 \cup X_2 \cup \{ 4,5,6,7,8,9\}$.
This case is symmetric to Case 2.

\item[\textbf{Case 4:}] $\{u,v\} \subseteq X_0 \cup X_2 \cup \{ 4,5,6,7,8,9\}$.
\begin{itemize}
  \item[\textbf{Case 4.1:}] $v \in X_2$.
  Then, for each~$x_4$ it follows that $\distP{x_4}\ge \hs a + a + \ell > \hs a + c+c'$.
   In order for $X_4\cup X_0 \cup \{8,9\}$ to not block,
   by the contra-positive of \Cref{obs:propPentagon4D},
   it follows that $X_0\subseteq \Pi(x_0)$ for all $x_0\in X_0$.
   This implies that $X_0 \cap \Pi(X_4)= \emptyset$ since $X_3 \subseteq \Pi(X_4)$.
   Hence, again, for $X_4 \cup X_0\cup \{8,9\}$ to not block,
   it must hold that $\Pi(X_0) = X_0 \cup X_1 \cup \{8,9\}$.
   Then, for each $x_1\in X_1$ it holds that
   $\distP{x_1}\ge \hs a + \ell + \ell > \hs a + c+c'$.
   Furthermore, since not all points in $X_2$ are matched together (since $v\in X_2$), by
   \Cref{obs:propPentagon4D},
   for each $x_2 \in X_2$ it holds that $\distP{x_2}\ge (\hs + 3) a  > \hs a + b+b'$.
   This implies that $X_2 \cup X_1 \cup \{2,3\}$ is blocking.

 \item[\textbf{Case 4.2:}] $v \notin X_2, u \in \{4,5\}$.
Then, for each $x_4 \in X_4$, it holds that $\distP{x_4} \geq \hs a + a + \ell' > \hs a + b+ b'$, 
where $\ell' = \dist(5, x_4) > \ell$.
The second inequality follows from the fact $a+\ell' > b+b'$.
In order for $X_0 \cup X_4 \cup \{8,9\}$ to not block,
by \Cref{lem:heptagon4DSM},
for each $x_0\in X_0$
 it holds that $X_0\subseteq \Pi(x_0)$
 such that
 $\distP{x_0} \leq \hs a + b+b'$. 
 In particular, it also implies that $X_0\cap \Pi(X_4)=\emptyset$.
 By the distance bound, it means that $\Pi(X_0) = X_0\cup X_1 \cup \{8,9\}$.
 Hence, for each agent~$x_2\in X_2$ it holds that
 $\distP{x_2}\ge \hs a + c + c'$.
 This implies that $X_2 \cup X_2\cup \{2,3\}$ is blocking.

 \item[\textbf{Case 4.3:}] $v \notin X_2, u \in \{6,7,8,9\}$. 
 Then, for each~$x_3 \in X_3$ it holds that $\distP{x_3} \ge \hs a + c + b > \hs a + b + b'$.
 In order for $X_2 \cup X_3 \cup \{4,5\}$ to not be blocking,
 by \Cref{obs:propPentagon4D}, 
 for each $x_2\in X_2$
 it holds that $X_2\subseteq \Pi(x_2)$
 such that $\distP{x_2}\le \hs a + c + c'$.
 Moreover, it must hold that $X_1 \subseteq \Pi(X_2)$.
 Hence, for each $x_1\in X_1$ it holds that
 $\distP{x_1}\ge \hs a + c + c'$.
 In order for $X_0 \cup X_1 \cup \{0,1\}$ to not block,
 by \Cref{obs:propPentagon4D}, 
 for each $x_0\in X_0$
 it holds that $X_0\subseteq \Pi(x_0)$
 such that
 $\distP{x_0}\le \hs a + c + c'$.
 This implies that at least $\hs$ agents from $X_1\cup X_4$ must be matched with $X_0$, a contradiction. 

  \item[\textbf{Case 4.5:}] $v \notin X_2, u \in X_0$. This case is symmetric to Case 2.1.

\end{itemize}
\end{itemize}

Therefore, in each case we get a contradiction showing that no agents~$u$ and $v$ exist which
can be matched with $X_3 \cup X_4$ in a \dsm. Hence,
no \dsm{} exists for which $\Pi(0)\cap Y =\emptyset$ and $\Pi(1)\cap Y=\emptyset$. 

\medskip
\noindent \textbf{The case when $\di$ is odd.} Suppose, towards a contradiction,
that $\Pi(0)\neq Y\cup \{0\}$.
\newH{Then, we infer that $\Pi(0) \cap Y = \emptyset$ since otherwise $Y\cup \{0\}$ is blocking due to the following:
  By construction, $Y\cup \{0\}$ is the unique most preferred coalition of agent~$0$.
  Moreover, by assumption, at most $\di-2$ points from $Y$ are matched together; recall that $|Y|=\di-1$.
  Then, $\distP{y'} \geq 2(b / 2)   >  b-\varepsilon$ for each $y' \in Y$; recall that all remaining points are at distance at least~$b/2$ to~$Y$.  
  Hence, $Y\cup \{0\}$ is blocking.
}

To improve readability, in the following, the subscript~$i+1$ in $X_{i+1}$ is taken modulo $5$.
  Then the most preferred coalition for $0$ which does not intersect~$Y$ is $X_0 \cup X_1 \cup \{0\}$. 
  The proof will be similar to the proof of \Cref{lem:heptagon4DSM}.
  First, we show that there exists $i \in \{0, \ldots, 4\}$ such that $X_i \cup X_{i +1} \subseteq \Pi(x)$ for some $x \in X_i$. That is at least one ``pentagon edge'' is matched together.
  Similar to the case for $\di$ being even, we begin with a simple observation

\begin{claim}\label{obs:proppentagon_nD}
For each $i \in \{0, \ldots, 4\}$ 
it holds that if $X_i \nsubseteq \Pi(x_i)$ for some $x_i\in X_i$, then $\distP{x_i} \geq (\hs+2)a$ for each~$x_i\in X_i$. %
\end{claim}
\begin{proof}[Proof of \cref{obs:proppentagon_nD}]
  \renewcommand{\qedsymbol}{$\diamond$}
If $X_i \nsubseteq \Pi(x_i)$, then $|\Pi(x_i) \cap X_i| \leq \hs-1$ since $|X_i| = \hs$. Then at least $\hs+2$ points in $\Pi(x_i)$ are not in $X_i$. Hence, $\distP{x_i} \geq (\hs+2)a$.
\end{proof}

Whenever $X_i \subseteq \Pi(x_i)$ for some $x_i \in X_i$, we abuse the notation to write the matching coalition containing $X_i$ as $\Pi(X_i)$ for $i \in \{0, \ldots, 4\}$.
Towards contradiction to our claim that at least one pentagon edge is matched, suppose that for each $i\in \{0,\ldots, 4\}$,
there exists a point~$x\in X_i$ such that that~$X_i \cup X_{i +1} \nsubseteq \Pi(x)$.
Without loss of generality, we consider two cases based on how $X_3$ is matched. We will show contradiction in both cases, thereby implying $X_i \cup X_{i +1 \bmod 5} \subseteq \Pi(x)$ for some $x \in X_i$ for some $i \in \{0, \ldots, 4\}$.

\begin{itemize}
  \item[\textbf{Case 1:}] $X_3 \nsubseteq \Pi(x_3)$ for some $x_3 \in X_3$.
  Then by \Cref{obs:proppentagon_nD}, $\distP{x_3} \geq (\hs+2)a > \hs a + c$.
  In order for $X_3\cup X_4 \cup \{3\}$ and $X_3\cup X_2\cup \{2\}$ to not block, respectively,
  by \cref{obs:proppentagon_nD},
  it must hold that $X_4\subseteq \Pi(x_4)$ and $\distP{x_4}\le \hs a + b$ for all~$x_4\in X_4$,
  and $X_2\subseteq \Pi(x_2)$ and $\distP{x_2}\le \hs a + c$ for all~$x_2\in X_2$.
  In other words, at least $\hs$ agents from $X_3\cup X_0$ have to be matched with $X_4$
and at least $\hs$ agents from $X_3\cup X_1$ have to be matched with $X_2$.
Since $X_3 \nsubseteq \Pi(x_3)$ for some $x_3 \in X_3$,
it follows that $X_0 \cap \Pi(X_4) \neq \emptyset$ and $X_1 \cap \Pi(X_2) \neq \emptyset$.
We distinguish between two cases.

If $X_0\subseteq\Pi(x_0)$ for some $x_0\in X_0$,
then $X_0 \subseteq \Pi(X_4)$.
By the distance bound for the agents in $X_4$,
it follows that the last agent in $\Pi(X_4)\setminus (X_4\cup X_0)$ is either $3$ or some agent from $X_3$.

In both cases,
it follows that $\distP{x_0}\ge \hs a + \ell > \hs a + c$.
In order for $X_0\cup X_1\cup \{0\}$ to not block (recall that $0\cup Y\notin \Pi$),
by \cref{obs:proppentagon_nD},
it follows that $X_1\subseteq \Pi(x_1)$ so that $\distP{x_1} \le \hs a + b$ for all $x_1\in X_1$.
Since no agent in $X_0$ is matched to $X_1$,
we have that $\Pi(x_1)= X_2\cup X_1\cup \{0\}$.
This implies that $X_2\cup X_3\cup \{2\}$ is blocking since $X_3\nsubseteq \Pi(x_3)$ for some $x_3\in X_3$.

\item[\textbf{Case 2:}] $X_3 \subseteq \Pi(x_3)$ for some $x_3 \in X_3$, i.e, all points in~$X_3$ are matched together.
Since no pentagon edge is matched among each other, meaning that $X_2\nsubseteq \Pi(X_3)$.
\begin{itemize}
\item[\textbf{Case 2.1:}]
If $X_4\subseteq \Pi(X_4)$ for some $x_4\in X_4$,
then no point in~$X_4$ is matched with $X_3$
and at most $\hs-1$ points from $X_0$ is matched with $X_4$; recall that we assume that no pentagon edge is matched to each other.
This implies that
$\distP{x_4}\ge (\hs-1)a + b + c > \dist(x_3, \{3\}\cup X_3\cup X_4)$ for all $x_4\in X_4$.
Furthermore, at most $\hs-1$ points from $X_2$ can be matched with $X_3$,
implying that
$\distP{x_3}\ge (\hs-1)a + b + c > \dist(x_3, \{3\}\cup X_4\cup X_3)$ for all $x_3\in X_3$.
This results in $X_3\cup X_4\cup \{3\}$ blocking~$\Pi$.

\item[\textbf{Case 2.2:}] If $X_4\nsubseteq \Pi(x_4)$ for some $x_4\in X_4$, then $\distP{x_4} \geq (\hs +2)a > \hs a + c$ by \cref{obs:proppentagon_nD}. This case is symmetric to Case 1 by replacing $X_3$ with $X_4$ as shown next.

In order for $X_0 \cup X_4 \cup \{4\}$ to not block~$\Pi$, there must exist an~$x_0 \in X_0$ with $\distP{x_0} \leq \hs a + b < (\hs +2)a$ since $b < 2a$.
By \cref{obs:proppentagon_nD}, we have that~$X_0 \subseteq \Pi(x_0)$.
By assumption, we further infer that $\Pi(X_0) \cap X_1 \neq \emptyset$.

But since no ``pentagon edge'' is matched, it follows that $X_1 \nsubseteq \Pi(X_0)$. Therefore, by \cref{obs:proppentagon_nD}, we have that $\distP{x_1} \geq (\hs+2)a$ for each $x_1 \in X_1$. Then, in order for coalition~$X_1 \cup X_2 \cup \{1\}$ to not be blocking, there must exist a point~$x_2 \in X_2$ with $\distP{x_2} \leq \dist(x_2, X_1 \cup X_2 \cup \{1\}) = \hs a+b < ( \hs +2)a$. 
Therefore,  by \cref{obs:proppentagon_nD}, it follows that $X_2 \subseteq \Pi(x_2)$ for each~$x_2 \in X_2$, and $\Pi(X_2) \cap X_3 \neq \emptyset$ since not every point in~$X_1$ can be matched to all points in~$X_2$.
Recall that  we assumed in this case that $X_3 \subseteq \Pi(x_3)$ for some $x_3 \in X_3$. Then $X_2 \cup X_3$ are matched together, contradicting the assumption that no ``pentagon edge'' is matched together.
\end{itemize}

\end{itemize}

Hence, we proved that $X_i \cup X_{i+1} $ are matched together for some $i \in \{0, \ldots, 4\}$.
Without loss of generality,  let $X_3 \cup  X_4 \cup \{u\}\in \Pi$. 
As in the proof of \Cref{lem:pentagon}, for each possible value of~$u$ we will show a contradiction  to stability of $\Pi$, showing that no agent~$u$ exists which is matched with $X_3 \cup X_4$ in a stable matching. 

\begin{itemize}
\item[\textbf{Case 1:}] $u \notin X_0 \cup X_2 \cup \{2,3\}$.
Then, $ X_3 \cup X_4 \cup \{3\}$ is blocking.

\item[\textbf{Case 2:}] $u \in X_0\cup \{3\}$.
Then, $\distP{x_3}\geq \hs a +c > \hs a +b$ for each $x_3 \in X_3$.
In order for $X_2 \cup X_3 \cup\{2\}$ to not be blocking,
by \cref{obs:proppentagon_nD},
it follows that
$X_2\subseteq \Pi(x_2)$ such that $\distP{x_2}\le \hs a + c$ for all $x_2\in X_2$.
In particular, this means that $X_2$ and $X_1$ are matched together such that
$2\in \Pi(X_2)$ or $1\in \Pi(X_2)$.
In any case, $\distP{x_1} \ge \hs a + c > \hs a + b$ for all $x_1\in X_1$.
In order for $X_0 \cup X_1 \cup\{0\}$ to not be blocking (recall that $\Pi(0)\cap Y = \emptyset$),
by \cref{obs:proppentagon_nD},
it follows that
$\distP{x_0}\le \hs a + c$ for all $x_0\in X_0$,
implying that at least $\hs$ points from $X_4\cup X_1$ have to be matched with $X_0$.
However, this is impossible since
$X_3\cup X_4 \cup \subseteq \Pi(X_4)$
and $X_1\cup X_2\subseteq \Pi(X_1)$.

\item[\textbf{Case 3:}] $u \in X_2\cup \{2\}$.
Then, $\distP{x_4}\geq \hs a +d >\hs a +c$ for each $x_4 \in X_4$.
In order for $X_0\cup X_4\cup \{4\}$ to not be blocking,
by \cref{obs:proppentagon_nD},
it follows that
$X_0\subseteq \Pi(x_0)$ such that $\distP{x_0}\le \hs a + b$ for all $x_0\in X_0$.
This implies that $\Pi(X_0)=X_0\cup X_1\cup \{4\}$.
Consequently,
we have that
$\distP{x_1}\ > \hs a + \ell > \hs a + c$ for all $x_1\in X_1$.
In order for $X_1\cup X_2\cup \{1\}$ to not be blocking,
by \cref{obs:proppentagon_nD},
it follows that
$\distP{x_2}\le \hs a + b$ for all $x_2\in X_2$.
However, this is impossible since $X_3 \cup X_4 \subseteq \Pi(X_3)$
and $X_1 \cup X_0 \subseteq \Pi(X_1)$.

\end{itemize}

Therefore, in each of case we get a contradiction showing  that no agent~$u$ exists which is matched with $X_3 \cup X_4$ in a stable matching. Hence,
no \dsm{} exists that does not contain $Y \cup \{0\}$.
\end{proof}
}

\begin{figure}[t!]
  \captionsetup[subfigure]{justification=centering}
  \centering
  \begin{subfigure}[t]{.45\textwidth}
    \begin{tikzpicture}[scale=1.4,every node/.style={scale=0.85}]
  \node[nn] at (0,0) (ui) {};
  \node[left = 0pt of ui] {$u_i$};
  \node at (1.5,-1) (s1) {};
  \node at (0,-1) (s3) {};
  \node at (0,1) (s2) {};

  \node[right=0pt of s1] {$w_j$};
  \node[left=0pt of s2] {$w_k$};
  \node[left=0pt of s3] {$w_{\setind}$};

  \path[draw] (ui) -| (s1);
  \path[draw] (ui) -- (s2);
  \path[draw] (ui) -- (s3);

  \begin{scope}[shift={(2.5,0)}]
    \node[nn] at (0:0.5) (s1) {};
    \node[nn] at (120:0.5) (s2) {};
    \node[nn] at (240:0.5) (s3) {};
    \node[nnn] at (0:0) (i) {};

    \node[above=0pt of s1] {$u_i^j$};
    \node[left= 0pt of s2] {$u_i^k$};
    \node[left=0pt of s3] {$u_i^{\setind}$};

    \node[above right=-3pt and -4pt of i] {$U_i$};

    \foreach \i / \j in {i/s1,i/s2,i/s3,s1/s2,s2/s3,s3/s1} {
      \path[ele] (\i) -- (\j);
    }

    \begin{pgfonlayer}{background}
      \begin{scope}[shift={(s2)}]
        \path[gl] (0,0) -- (0,0.5);
      \end{scope}
      
      \begin{scope}[shift={(s3)}]
        \path[gl] (0,0) -- (0,-0.5);
      \end{scope}

      \begin{scope}[shift={(s1)}]
        \path[gl] (0,0) -- (.8,0);
        \path[gl] (.8,0) -- (.8,-1);
      \end{scope}
    \end{pgfonlayer}
    
  \end{scope}
\end{tikzpicture}
\caption{Gadget (right) for an element vertex~$u_i$ (left) s.t.\  element~$i$ belongs to sets~$S_j,S_k,S_{\setind}$.}\label{fig:elt-gadG}
\end{subfigure}~~~\begin{subfigure}[t]{0.45\linewidth}
  \begin{tikzpicture}[scale=1.4,every node/.style={scale=0.75}]
  \node[nn] at (0,0) (ui) {};
  \node[below = 0pt of ui] {$w_j$};
  \node at (-1,0) (s1) {};
  \node at (1,0) (s3) {};
  \node at (0,1) (s2) {};

  \node[below=0pt of s1] {$u_p$};
  \node[left=0pt of s2] {$u_i$};
  \node[below=0pt of s3] {$u_q$};

  \path[gl] (ui) -- (s1);
  \path[gl] (ui) -- (s2);
  \path[gl] (ui) -- (s3);

  \begin{scope}[shift={(2:2.5)}]
    \node[nn] at (-30:0.5) (s1) {};
    \node[nn] at (90 :0.5) (s2) {};
    \node[nn] at (210:0.5) (s3) {};

    \node[sss] at (0:0) (cent) {};
    
    \node[below=0pt of s1] {$w_j^p$};
    \node[left= 0pt of s2] {$w_j^i$};
    \node[below =0pt of s3] {$w_j^{q}$};
    \node[below right = -2pt and -1pt of cent] {$W_j$};

    \foreach \i / \j in {s1/s2,s2/s3,s3/s1} {
      \path[sett] (\i) -- (\j);
    }
    
    \begin{pgfonlayer}{background}
      \begin{scope}[shift={(s2)}]
        \path[gl] (0,0) -- (0,0.3);
      \end{scope}

      \begin{scope}[shift={(s3)}]
        \path[gl] (0,0) -- (1.5,0);
      \end{scope}
      
      \begin{scope}[shift={(s1)}]
        \path[gl] (0,0) -- (-1.5,0);
      \end{scope}
    \end{pgfonlayer}
  \end{scope}
\end{tikzpicture}
\caption{Gadget (right) for a set-vertex~$w_j^i$ for which the set~$S_j$ consists of three elements~$i,p,q$.}\label{fig:set-gadG}
\end{subfigure}
\caption{Element and set gadgets described in \cref{sub:construction-G}.}\label{fig:gadget-R}
\end{figure}
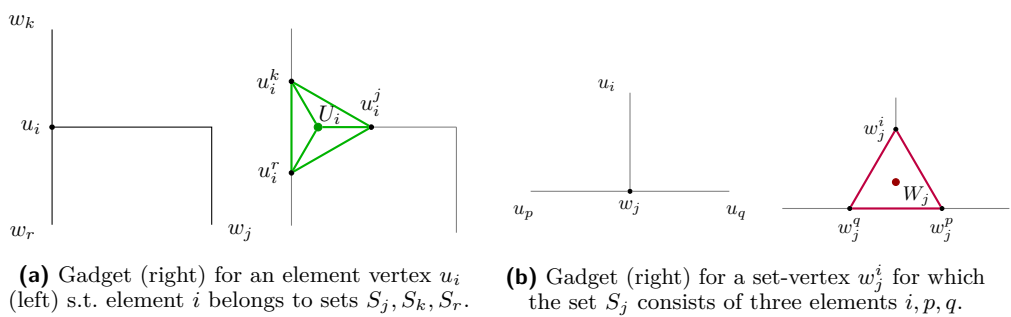

\subparagraph{The remaining gadgets.} %
Let $I=(X, \mathcal{S})$ be an instance of \xcts{}.
Similarly to the case with $\di=3$, we first embed the associated graph~$G(I)=(U\cup W, E)$ into a $2$-dimensional grid with edges drawn using line segments of length at least $L\ge 200$, and with parallel lines at least $4L$ grid squares apart.
The element- and the edge-gadget are almost the same as the ones describe in \cref{sub:element-set-gadgets,sub:enforcement}.
The only difference is that we replace each element-agent~$u_i$ (for $u_i \in U$) with a size-($\di - 2$) coalition~$U_i$ that are embedded so close to each other that any stable matching must match them together.
Similarly, for each $z\in [\enn]$ (recall that $\enn$ is a constant as defined in the \cref{sub:3-sr-construction}) and $w_j \in W$, we replace the two agents~$\alpha_i^j[z]$ and $\beta_i^j[z]$ with a size-($\di-1$) coalition~$\hat{A}_i^j[z]$ such that the distance between each pair of points in $\hat{A}_i^j[z]$ is close to zero, and define \myemph{$A_i^j[z]$} $\coloneqq \hat{A}^j_i[z]\cup \{\gamma_i^j[z]\}$.
For each set-vertex~$w_j\in W$, assume that the three connecting edges in~$G(I)$ are going rightward, leftward, and upward, connecting the element-vertices~$u_i$, $u_p$, $u_q$, respectively.
We create three set-agents, called $w_j^i$, $w_j^p$, $w_j^q$, and an additional coalition~$W_j$ of size $\di-3$ and as before, define $w_j^i=\gamma^i_j[\enn]$.
We embed them into~$\mathds{R}^2$ in such a way that $w_j^i,w_j^p,w_j^{q}$ are on the segment of the rightward, leftward, and upward edge, respectively,
and are of {equidistance~\settc{$17.5$}} to each other,
and the coalition~$W_j$ is embedded in the center so that the distance between any two of them is close to zero.
Moreover, the largest distance from any agent of $W_j$ to any agent of $\{w_j^i, w_j^p, w_j^q\}$ is \settc{$10$}. %
See Figure~\ref{fig:gadget-R} for an illustration.

We remark that by the construction of the set-gadget and the edge-gadget,
each set-agent~$w^i_j$ prefers coalition~$A^j_i[{\enn}]$ (recall that $\gamma_i^j[z]=w^i_j$) to coalition~$\{w^i_j,w^p_j,w^q_j\}\cup W_j$ since the sum of distances from $w^i_j$ to the latter coalition is {$17.5+17.5+10(\di-3) > (\di-1)\cdot (10-\varepsilon)$}.
To ensure that one of the two coalitions is chosen, we make use of the star-gadgets from \cref{ex:heptagon_n-DSM}.
Define $b\coloneqq 22.6$ and $c\coloneqq 22.7$.
We create an agent-subset~$F_j^i$ of size~$\di-1$ and agent~$h_j^i$ and a star-gadget~$W$ as described in \cref{ex:heptagon_n-DSM}, with $Y$ being the extra $\di-1$ agents such that the most preferred coalition of each agent in~$Y$ is {$Y\cup \{h_j^i\}$}.
Note that $F_j^i$ has the same role as $\{f_j^i, g_j^i\}$ in the case for $\di=3$.
\begin{itemize}%
  \item
The distance between each two agents in $F_j^i$ is close to zero.
 \item
The distance from each agent in~$F^i_j$ to each agent in~$\hat{A}^j_i[{\enn}]$ is in the range of $[10+\varepsilon, 10+2\varepsilon)$.
 \item
The distance from each agent in $F^i_j$ to agent~$w_j^i$ is {$10+\frac{15}{\di-1}$}.
 \item
The distance from each agent in~$F^i_j$ to agent~$h^i_j$ is $15+2\varepsilon$.
 \item
The distance from agent~$h^i_j$ and each agent~$Y$ is $15+ 3\varepsilon$.
 \item
The distance from each agent~$Y$ to $0$ (and also to $1$ if $\di$ is even) is $15+ 4\varepsilon$. 
\end{itemize}

\begin{figure}[t!]
  \centering
\begin{tikzpicture}[scale=1.2, every node/.style={scale=0.8}]
  \def\d{.54}
  \def\f{.6}
  \def\e{.3}
  \def\k{.5}
  
  \node[nn] at (0:\e) (s1) {};
  \node[nn] at (120:\e) (s2) {};
  \node[nn] at (240:\e) (s3) {};
  \node[nnn] at (0:0) (i) {};
  
  \node[above=-1pt of s1] {$u_i^j$};

  \foreach \i / \j in {i/s1,i/s2,i/s3,s1/s2,s2/s3,s3/s1} {
    \path[ele] (\i) -- (\j);
  }
 
  \node[below right=1pt and -1pt of i, inner sep=.2pt, fill=white] {\scriptsize $U_i$};

  \begin{pgfonlayer}{background}
  \begin{scope}[shift={(s2)}]
    \path[gl] (0,0) -- (0,0.5);
  \end{scope}
  
  \begin{scope}[shift={(s3)}]
    \path[gl] (0,0) -- (0,-0.5);
  \end{scope}
  \end{pgfonlayer}
    \coordinate (x) at (s1);
    \begin{scope}[shift={(0.8,0)}]
      \node[dnode] at (0,0.03) (f1) {};
      \node[dnode] at (0,-0.03) (f2) {};
      \node[nn] at (\k,0) (f3) {};
      \begin{pgfonlayer}{background}
        \draw[gl] (x) -- (f1) -- (f3);
        \draw[gl] (x) -- (f2) -- (f3);
      \end{pgfonlayer}
      \node[above = -0.5pt of f1] {\small $\hat{A}^{j}_i[1]$};
      \node[above = 0.5pt of f3] {\small $\gamma^j_i[1]$};
    \end{scope}

    \coordinate (x) at (f3);
    \def\st{1.3}
    
     \foreach \i/\j/\k/\l in {1/2.01/3/4,3.03/4.06/5/6,
      5.1/6.15/7/8,7.21/8.28/9/10} {
      \begin{scope}[shift={(\st,0)}]
        \node[dnode] at (\i*\d,0.03) (f1) {};
        \node[dnode] at (\i*\d,-0.03) (f2) {};
        \node[nn] at (\j*\d,0) (f3) {};
        \begin{pgfonlayer}{background}
 \draw[gl] (x) -- (f1) -- node[text=gray, above, inner sep=0pt] (f13) {\small $\overbrace{~~~~~~~}$} (f3);
          \node[above = -0.5pt of f13] {\scriptsize $8\!+\!\varepsilon_{\l}$};
          \draw[gl] (x) -- node[text=gray, below, inner sep=0pt] (f23) {\small $\underbrace{~~~~~~~}$} (f2) -- (f3);
          \node[below = -0.5pt of f23] {\scriptsize $8\!+\!\varepsilon_{\k}$};
        \end{pgfonlayer}
      \end{scope}
      \coordinate (x) at (f3);
    }
    
      \foreach \i/\j/\k/\l in {9.36/10.45/11/12} {
      \begin{scope}[shift={(\st,0)}]
        \node[dnode] at (\i*\d,0.03) (f1) {};
        \node[dnode] at (\i*\d,-0.03) (f2) {};
        \node[nn] at (\j*\d,0) (f3) {};
        \begin{pgfonlayer}{background}
          \draw[gl] (x) -- (f1) --  (f3);
          \draw[gl] (x) --  (f2) -- (f3);
        \end{pgfonlayer}
      \end{scope}
      \coordinate (x) at (f3);
	}
  \node[above = 1pt of f1] {\small $\hat{A}^{j}_i[z']$};
  \node[above = 1.5pt of f3] {\small $\gamma^j_i[z']$};
  \node[draw=none, right = 5ex of f3] (bendingpoint) {};
  \node[right = -6pt of bendingpoint] {\small bending point};
  
  \draw[->, shorten <= 1pt, shorten >= 1pt] (bendingpoint) edge[bend right = 10] (f3);
  
  \begin{scope}[shift={(x)}]
    \foreach \i/\j in {-1.1/-2.21,-3.33/-4.46,-5.6/-6.75,-7.91/-9.08,-10.26/-11.45} {
      \begin{scope}
        \node[dnode] at (-0.03,\i*\d) (f1) {};
        \node[dnode] at (0.03,\i*\d) (f2) {};
        \node[nn] at (0,\j*\d) (f3) {};
        \begin{pgfonlayer}{background}
          \draw[gl] (x) -- (f1) -- (f3);
          \draw[gl] (x) -- (f2) -- (f3);
        \end{pgfonlayer}
      \end{scope}
      \coordinate (x) at (f3);
    }
  \end{scope}
  \coordinate (an) at (f1);
  \coordinate (bn) at (f2);

  \node[above right = -2pt and  -2pt of f2] {\small $\hat{A}^j_i[{\enn}]$};
  \begin{scope}[shift={(x)}]
    \node[nn] at (x) (s1) {};
    \node[nn] at (-60:\f) (s2) {};
    \node[nn] at (240:\f) (s3) {};
    \node[nn] at (0,-\e) (s4) {};

    \node[right=0pt of s1] {\small $w_j^i=\gamma^j_i[{\enn}]$};
    \node[below left = 0pt and -5pt of s2] {\small $w_j^p$};
    \node[below right = 0pt and -5pt of s3]  {\small $w_j^{q}$};    
    \node[below right = -2pt and -5pt of s4]  {\tiny $W_j$};

    \foreach \i / \j in {s1/s2,s2/s3,s3/s1} {
      \path[sett] (\i) -- (\j);
    }
    \path[sett] (s1) -- node[text=black,left, fill=white, inner sep=1pt,xshift=3pt] {\tiny $17.5$} (s3);
  \end{scope}

  \coordinate (x) at (s3);
  \begin{scope}[shift={(x)}]
    \foreach \i / \j in {-1.09/-2.17} {
      \begin{scope}
        \node[dnode] at (\i*\d,-0.03) (f1) {};
        \node[dnode] at (\i*\d,0.03) (f2) {};
        \node[nn] at (\j*\d,0) (f3) {};
        \begin{pgfonlayer}{background}
          \draw[gl] (x) -- (f1) -- (f3);
          \draw[gl] (x) -- (f2) -- (f3);
        \end{pgfonlayer}
      \end{scope}
      \coordinate (x) at (f3);
    }

    \coordinate (aqn) at (f1);
    \coordinate (bqn) at (f2);
    
    \foreach \i/\j in {-3.24/-4.3,-5.35/-6.39} {
      \begin{scope}
        \node[dnode] at (\i*\d,-0.03) (f1) {};
        \node[dnode] at (\i*\d,0.03) (f2) {};
        \node[nn] at (\j*\d,0) (f3) {};
        \begin{pgfonlayer}{background}
          \draw[gl] (x) -- (f1) -- (f3);
          \draw[gl] (x) -- (f2) -- (f3);
        \end{pgfonlayer}
      \end{scope}
      \coordinate (x) at (f3);
    }

    \foreach \i in {-7,-7.5,-8} {
        \node[nn] at (\i*\d,0) {};
    }
    \end{scope}    
    
    \coordinate (x) at (s2);
    \begin{scope}[shift={(x)}]
      \foreach \i / \j in {1.09/2.17} {
      \begin{scope}
        \node[dnode] at (\i*\d,-0.03) (f1) {};
        \node[dnode] at (\i*\d,0.03) (f2) {};
        \node[nn] at (\j*\d,0) (f3) {};
        \begin{pgfonlayer}{background}
          \draw[gl] (x) -- (f1) -- (f3);
          \draw[gl] (x) -- (f2) -- (f3);
        \end{pgfonlayer}
      \end{scope}
      \coordinate (x) at (f3);
    }
    
    \coordinate (apn) at (f1);
    \coordinate (bpn) at (f2);
    
   \foreach \i/\j in {3.24/4.3,5.35/6.39} {
      \begin{scope}
        \node[dnode] at (\i*\d,-0.03) (f1) {};
        \node[dnode] at (\i*\d,0.03) (f2) {};
        \node[nn] at (\j*\d,0) (f3) {};
        \begin{pgfonlayer}{background}
          \draw[gl] (x) -- (f1) -- (f3);
          \draw[gl] (x) -- (f2) -- (f3);
        \end{pgfonlayer}
      \end{scope}
      \coordinate (x) at (f3);
    }

    \foreach \i in {7,7.5,8} {
      \node[nn] at (\i*\d,0) {};
    }

    \end{scope}

    \def\f{0.7}

    \node[] at ([shift={(150:\d)}]s1) (x) {};
    
    \begin{scope}[shift={(x)},rotate=-40]
      \foreach \i in {0} {
        \begin{scope}[shift={(\i*\f,0)}]
          \node[dnode] at (0,-0.03) (f1) {};
          \node[dnode] at (0,0.03) (f2) {};
          \node[nn] at (-\f,0) (f3) {};

          \begin{pgfonlayer}{background}
            \foreach \s in {f1,f2} {
              \foreach \t in {s1,an,bn}
              {
                \draw[ml] (\s) --  (\t);
              }
            }
            \draw[ml] (f2) -- node[text=black, left, fill=white, inner sep=1pt,xshift=5.5pt,yshift=5pt,rotate=45] {\tiny $10\!+\!{15\over \di-1}$} (s1);
            \draw[pl] (f2) -- node[text=black, above, fill=white, inner sep=0.5pt,yshift=-1pt,rotate=30] {\scriptsize $15\!+\varepsilon$} (f3);
            \draw[pl] (f1) -- node[text=black, above, fill=white, inner sep=0.5pt,rotate=30] {\scriptsize $15\!+\!2\varepsilon$} (f3);
          \end{pgfonlayer}
        \end{scope}
        \coordinate (x) at (f3);
      }
      \node[below left = 0pt and 0pt of f2] {\small $F^i_j$};
      \node[below left = -2pt and -2pt of f3] {\small $h^i_j$};
      
      \foreach \i in {-2} {
        \begin{scope}[shift={(\i*\f,0)}]
          \node[dnode] at (0,-0.03) (f1) {};
          \node[dnode] at (0,0.03) (f2) {};
          \node[nn] at (-\f,0) (f3) {};
          \begin{pgfonlayer}{background}
            \draw[pl] (x) -- (f1) -- (f3);
            \draw[pl] (x) -- node[text=black, above, fill=white, inner sep=0.5pt,yshift=-1pt,rotate=30] {\scriptsize $15\!+\!3\varepsilon$} (f2) -- node[text=black, above, fill=white, inner sep=0.5pt,yshift=-5pt,rotate=30] {\scriptsize $15\!+\!4\varepsilon$} (f3);
          \end{pgfonlayer}
        \end{scope}
        \coordinate (x) at (f3);
      }
      \node[above right = 0pt and -5pt of f1] {};
      \node[below left = 0pt and 0pt of f2] {$Y$};
    \end{scope}

    \coordinate (x) at (f3);
    
    \begin{scope}[scale=0.072,shift={(x)}]
      \def\xx{1}
      \def\yy{1}
      \def\degr{10}
      \def\y{1}
      \def\nd{10}
      \def\a{6.6}
      \def\b{10.1}
      \def\c{10.2}
      \def\bb{\b*\b}
      \def\aa{\a*\a}
      \def\cc{\c*\c}

      \pgfmathsetmacro\xx{\a/sin(36)/2}
      
      \pgfmathsetmacro\degr{54+acos((\bb + \aa - \cc) / (2*\a*\b) )}
      \pgfmathsetmacro\yy{\bb+\xx*\xx-2*\b*\xx*cos(\degr)}
      
      \pgfmathsetmacro\y{sqrt(\yy)}
      \pgfmathsetmacro\nd{acos((\xx*\xx+\yy-\bb)/(2*\xx*\y))}

      \begin{scope}
    	 \dpentevenN{6.6}{10.1}{10.2}{0}{10.1}{10.2}
    \dnamesneven{\a}{0}

      \end{scope}
    \end{scope}

     \node[] at ([shift={(-60:\d)}]s2) (x) {};
    
    \begin{scope}[shift={(x)},rotate=-60]
      \foreach \i in {0} {
        \begin{scope}[shift={(\i*\f,0)}]
          \node[dnode] at (0,-0.03) (f1) {};
          \node[dnode] at (0,0.03) (f2) {};
          \node[nn] at (\f,0) (f3) {};

          \begin{pgfonlayer}{background}
            \foreach \s in {f1,f2} {
              \foreach \t in {s2,apn,bpn}
              {
                \draw[ml] (\s) -- (\t);
              }
            }
            \draw[pl] (f1) -- (f3);
            \draw[pl] (f2) -- (f3);
          \end{pgfonlayer}
        \end{scope}
        \coordinate (x) at (f3);
      }
      
      \foreach \i in {2} {
        \begin{scope}[shift={(\i*\f,0)}]
          \node[dnode] at (0,-0.03) (f1) {};
          \node[dnode] at (0,0.03) (f2) {};
          \node[nn] at (\f,0) (f3) {};
          \begin{pgfonlayer}{background}
            \draw[pl] (x) -- (f1) -- (f3);
            \draw[pl] (x) -- (f2) -- (f3);
          \end{pgfonlayer}
        \end{scope}
        \coordinate (x) at (f3);
      }
    \end{scope}

    \coordinate (x) at (f3);

    \begin{scope}[scale=0.072,shift={(x)}]
      \def\xx{1}
      \def\yy{1}
      \def\degr{10}
      \def\y{1}
      \def\nd{10}
      \def\a{6.6}
      \def\b{10.1}
      \def\c{10.2}
      \def\bb{\b*\b}
      \def\aa{\a*\a}
      \def\cc{\c*\c}

      \pgfmathsetmacro\xx{\a/sin(36)/2}
      
      \pgfmathsetmacro\degr{54+acos((\bb + \aa - \cc) / (2*\a*\b) )}
      \pgfmathsetmacro\yy{\bb+\xx*\xx-2*\b*\xx*cos(\degr)}
      
      \pgfmathsetmacro\y{sqrt(\yy)}
      \pgfmathsetmacro\nd{acos((\xx*\xx+\yy-\bb)/(2*\xx*\y))}

      \begin{scope}
           \dpentevenN{\a}{\b}{\c}{150}{10.1}{10.2}
      \end{scope}
    \end{scope}

     \node[] at ([shift={(240:\d)}]s3) (x) {};
    
    \begin{scope}[shift={(x)},rotate=240]
      \foreach \i in {0} {
        \begin{scope}[shift={(\i*\f,0)}]
          \node[dnode] at (0,-0.03) (f1) {};
          \node[dnode] at (0,0.03) (f2) {};
          \node[nn] at (\f,0) (f3) {};

          \begin{pgfonlayer}{background}
            \foreach \s in {f1,f2} {
              \foreach \t in {s3,aqn,bqn}
              {
                \draw[ml] (\s) -- (\t);
              }
            }
            \draw[pl] (f1) -- (f3);
            \draw[pl] (f2) -- (f3);
          \end{pgfonlayer}
        \end{scope}
        \coordinate (x) at (f3);
      }
      
      \foreach \i in {2} {
        \begin{scope}[shift={(\i*\f,0)}]
          \node[dnode] at (0,-0.03) (f1) {};
          \node[dnode] at (0,0.03) (f2) {};
          \node[nn] at (\f,0) (f3) {};
          \begin{pgfonlayer}{background}
            \draw[pl] (x) -- (f1) -- (f3);
            \draw[pl] (x) -- (f2) -- (f3);
          \end{pgfonlayer}
        \end{scope}
        \coordinate (x) at (f3);
      }
    \end{scope}

    \coordinate (x) at (f3);

    \begin{scope}[scale=0.072,shift={(x)}]
      \def\xx{1}
      \def\yy{1}
      \def\degr{10}
      \def\y{1}
      \def\nd{10}
      \def\a{6.6}
      \def\b{10.1}
      \def\c{10.2}
      \def\bb{\b*\b}
      \def\aa{\a*\a}
      \def\cc{\c*\c}

      \pgfmathsetmacro\xx{\a/sin(36)/2}
      
      \pgfmathsetmacro\degr{54+acos((\bb + \aa - \cc) / (2*\a*\b) )}
      \pgfmathsetmacro\yy{\bb+\xx*\xx-2*\b*\xx*cos(\degr)}
      
      \pgfmathsetmacro\y{sqrt(\yy)}
      \pgfmathsetmacro\nd{acos((\xx*\xx+\yy-\bb)/(2*\xx*\y))}

      \begin{scope}
          \dpentevenN{\a}{\b}{\c}{60}{10.1}{10.2}
      \end{scope}
    \end{scope}
  \end{tikzpicture}
  \caption{Gadget for edge~$\{u_i,w_j\}$ in $G(I)$ with $S_j=\{i,p,q\}$ for the case when $\di$ is even, omitting the garbage collector agents for the sake of brevity.}\label{fig:edge-gadget-G}
  \end{figure}
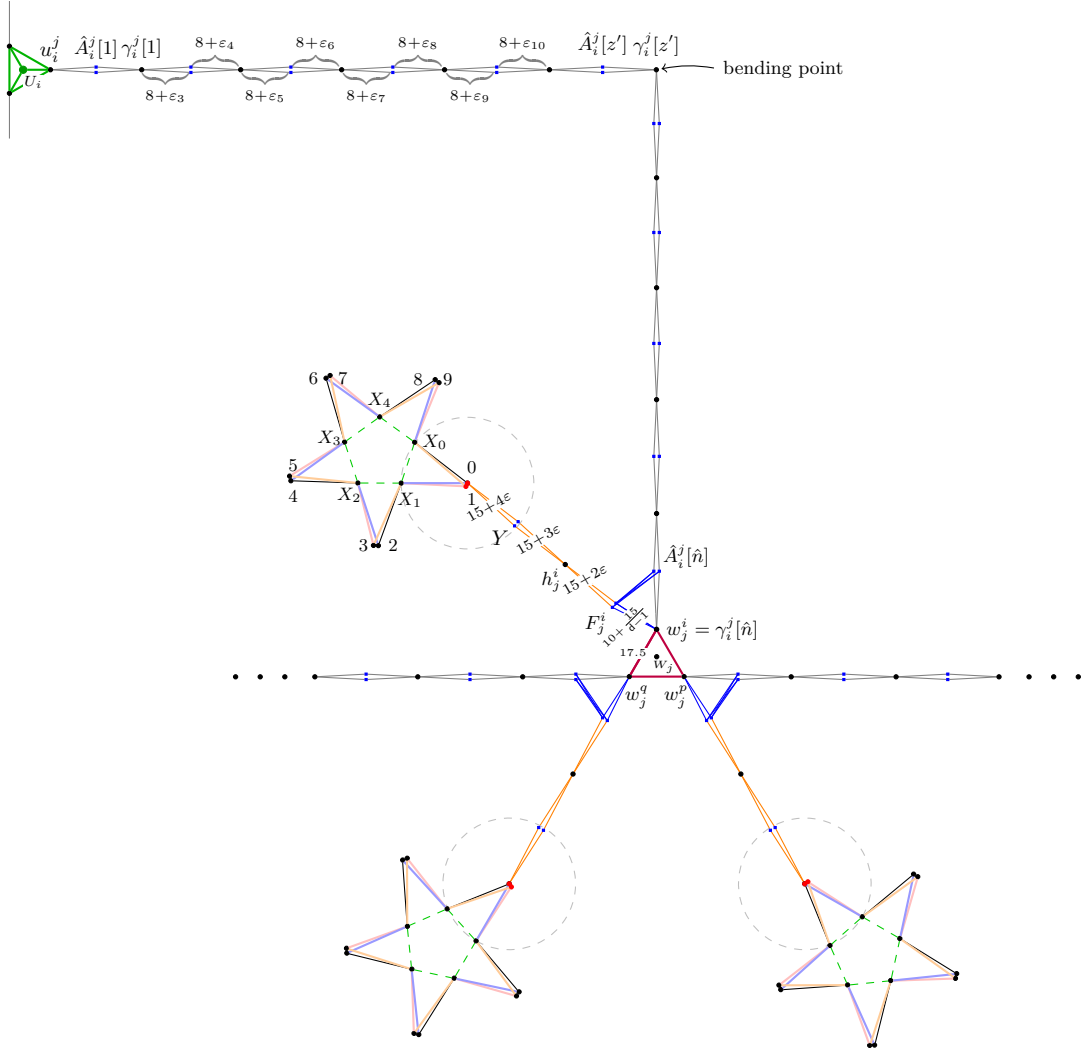

Finally, we create two types of garbage collector agents to match with some left over agents.
For each added star gadget corresponding to~$S_j$ and~$i\in S_j$, we create $O(\hs)$ garbage collector~ agents~$R_j^i$ as follows:
If $\di$ is odd, set $|R_j^i| \coloneqq \di - \hs - 2$.
Otherwise if $\di \le 6$, set $|R_j^i|\coloneqq 2\di - \hs - 5$, and otherwise set $|R_j^i|\coloneqq\di-\hs -5$.
These agents have distance close to zero to each other.
\newH{For each $y\in R_j^i$ it holds that
$\ell < \dist(y,x) < 2\ell < \dist(y,x')$,
where $x$ is an agent from the same star and $x'$ is an agent from neither $R_j^i$ or the same star.}
It is straightforward to see that the distance between any two agents from different star-gadgets
is larger than $\ell$,
and the distance from an agent in~$W$ to an agent to a set-gadget is at larger~$\ell$,
 where $a$, $b$, $b'$, $c'$, and $\ell$ are as defined in \Cref{ex:heptagon_n-DSM}.
 Lastly, we add $m-n$ triples of additional garbage collector agents.
 The agents in each triple have distance close to zero to each other but is far away from the other agents.
 Note that each triple will be matched to some~$W_j$ whenever $S_j$ is not chosen to the exact cover.
See \cref{fig:edge-gadget-G} (for even~$\di$, without the garbage collector agents) for an illustration.
This completes the description of the construction, which clearly can be done in polynomial time.

\appendixsection{sub:correctness-G}
\subsection{The correctness proof for $\di \ge 4$}\label{sub:correctness-G}
The reasoning for the correctness is similar to the one for~$\di=3$.
For the forward direction, assume that $(X,\mathcal{S})$ admits an exact cover~$\mathcal{K}$.
Then, using a reasoning similar to the one for $\di=3$, one can verify that the following \dm[\di]~$\Pi$ is stable; recall that $\hs=\lfloor (\di - 1) / 2\rfloor$. 
\begin{itemize}
  \item  For each $S_j\in \mathcal{K}$ with $S_j=\{i,p,q\}$ add $\{w_j^i,w_j^p,w^q_j\}\cup W_j$ to $\Pi$.
  \item For each element $i\in X$ let $S_{k}, S_{\setind}$ be the two sets which contain~$i$, but are not chosen in the exact cover~$\mathcal{K}$. Add $U_i \cup \{u_i^k,u_i^{\setind}\}$ to~$\Pi$.
  For each $S_j\notin \mathcal{K}$, take a triple of garbage collector agents (of the second type) and match them with~$W_j$.
  \item For each element~$i\in X$ and each set~$S_j\in\mathcal{S}$ with $i\in S_j$,
  \newH{call the agents in the associated star-gadget along with the tail $X_0\cup \cdots \cup X_4\cup \{0,1,2,3,4,h_j^i\} \cup Y \cup \{F_j^i\} \cup \{5,6,7,8,9\mid \text{ if } \di \text{ odd}\}$.
    If $S_j\in\mathcal{K}$, then add all $\hat{A}_j^i[z] \cup \{\gamma_j^i[z-1]\}$, $z\in [\enn]$, to $\Pi$.
    Otherwise, add all~$A^j_i[z]$, $z\in [\enn]$, to $\Pi$. 
    Add $F_{j}^i\cup \{h_j^i\}$ and $Y\cup \{0\}$ to~$\Pi$.}
    If $\di$ is odd, add $X_1\cup X_2\cup \{1\}$ and $X_3\cup X_4\cup \{3\}$ to~$\Pi$. %
    Otherwise, add $X_1\cup X_2\cup \{2,3\}$ and $X_3\cup X_4 \cup \{6,7\}$ to~$\Pi$. 
    \newH{Next, %
      if $\di \le 6$, then match $X_0$ with $\di-\hs$ agents from $(1,8,9,4)$ (in this sequence) to~$\Pi$.
      In any case, match the remaining star-agents with $R_j^i$.}      

\end{itemize}

\noindent The proof for the backward direction works analogously to~$\di=3$ and is deferred to the appendix. %
\label{page-correctnessG}
\appendixproofremainPart{thm:main}{the correctness}{$\di \ge 4$}{page-correctnessG}{
For the ``if'' direction, similar to \cref{lem:d=3-X3C->Stable}, we show that the $\Pi$ we defined is stable by showing no agent is in a blocking coalition.

  For each element $i\in X$ let $S_{k}, S_{\setind}$ be the two sets that contain~$i$, but are not chosen into the exact cover.
  Then, both $u_i^k$ and $u_i^{\setind}$ are matched to their most preferred coalition and do not form blocking. The same holds for $U_i$.

  Consider an arbitrary set~$S_j$ and let $i \in S_j$. A similar argument to \cref{lem:d=3-X3C->Stable} show that no agent from $A^j_i[z]$ is in a blocking coalition for each $z \in [\enn]$.

  Next, we show that $W_j$ is not involved in any blocking coalition.
  \newH{Clearly, if $S_j\in \mathcal{K}$, then each agent~$W_j$ is matched with its most preferred coalition and will not be blocking.
  If $S_j\notin \mathcal{K}$, then }
the agents~$W_j$ cannot form a blocking coalition with other agent in the set-gadget or the chain on the grid.
  We show that the agents~$W_j$ cannot form a blocking coalition with any star-gadget or its tail since by showing that no agent on a star-gadget or its tail is blocking later.
 
  Let us now consider the star-gadget and its tail corresponding to set~$S_j$ and element~$i$ with $i\in S_j$.
  We only show the case when $\di$ is even; the other case is similar. 
  \newH{No agent in $\{0,2,3,6,7, h_j^i\}$ is involved in a blocking coalition since
    each of them is matched to its most preferred coalition.
    Further, no agent~$x$ in~$X_2$ is involved in a blocking coalition since all coalitions~$T$ that $x$ prefers to~$\Pi(x)$ consist of $X_2$ and $\hs+2$ agents from $X_1\cup X_3$ but no agent from~$T\cap X_3$ will deviate.
    Similarly, no no agent in~$X_4$ is involved in a blocking coalition.
    Hence, no agent~$x$ in~$X_1$ is involved in a blocking coalition since all coalitions that it prefers to~$\Pi(x)$ involves at least one agent from~$\{0,2,3\}\cup X_2$. %
    Similarly, one can verify that no agent in~$X_0\cup \{8,9,1,4,5\}$ is involved in a blocking coalition.
  }
  
 Therefore, no coalition in the star-gadget is blocking.
 
 Next we claim that no agent from $Y\cup F^i_j$ is in a blocking coalition.
 $Y$ cannot be blocking since the only coalition preferred by $Y$ to $\Pi(Y)$ is $Y \cup\{h^i_j\}$.
 \newH{Similarly, no agent~$f$ in~$F_j^i$ is involved in a blocking coalition, since the coalition that $f$ prefers to~$\Pi(f)$ involves some agent from $A_i^j[\enn]$ but none of them is going to deviate. }
 Hence, $\Pi$ is stable.

Next we prove the ``only if'' direction of the correctness proof.

  Assume that $\Pi$ is a \dsm[\di]. 
We first prove a statement analogous to \cref{lem:set-agent}.
\begin{lemma}\label{lem:set-agentG}
  For each element~$i\in X$ and each set~$S_j$ with $S_j=\{i,p,q\}$ the stable matching~$\Pi$ satisfies the following. %
   \begin{enumerate}[(i)]
     \item\label{lem:fghG} $F_j^i\subseteq \Pi(h_j^i)$. %
     \item\label{lem:left-rightG} $\Pi$ contains either all coalitions~$A_i^j[z]$ or all coalitions~$A_i^j[z]\setminus \{\gamma_i^j[z]\} \cup \{\gamma^j_i[z-1]\}$,  $z\in [\enn]$.
   \end{enumerate}
\end{lemma}

{\begin{proof}
  The reasoning for odd~$\di$ is analogous to the case with $\di=3$; see the proof of \cref{lem:set-agent}.
  Hence, we assume that $\di$ is even with $\di = 2\hs + 2$.
  We drop $i$ and $j$ from the superscripts to improve readability.
  Furthermore, let $Y$ denote the extra agents from the star-gadget,
  $X_i$, $i\in \{0,\ldots,4\}$ denote the ``pentagon-agents'',
  and $0,\ldots,9$ denote the remaining agents in the gadget.
  First, by \cref{lem:heptagon4DSM}, we observe that there exists one agent~$x\in \{0,1\}$ with $\Pi(x)\cap Y \neq \emptyset$.

  Then, for each $y \in (Y\cap \Pi(0))\cup (Y\cap\Pi(1))$,
     we have that
     $\distP{y} \ge b-\varepsilon  > \dist(y,\{h\}\cup Y)$.
     For each $y\in Y\setminus (\Pi(0)\cup \Pi(1))$,
     we have that
     $\distP{y} > \dist(y, h) + \dist(y, 0) > \dist(y,\{h\}\cup Y)$ (note that besides the agents in~$Y$,
    agent~$h$ is the closest agent available to~$y$, followed by $0$).
   In order to prevent $Y\cup \{h\}$ from forming a blocking coalition,
   it must hold that $\dist(h, \Pi(h))\le \dist(h, Y)$.
   By construction, this implies that $\Pi(h)\cap F \neq \emptyset$.
   
   Now, we show the first statement.
   Suppose, for the sake of contradiction, that $F\nsubseteq \Pi(h)$.
   {This means that for each $f\in F\cap \Pi(h)$ it holds $\distP{f} \ge 15+2\varepsilon + 10+\varepsilon > \dist(f, F\cup \{h\})$.
     Further, for each $f\in F\setminus \Pi(h)$ it holds that $\distP{f} \ge 2(10+\varepsilon) > \dist(f, F\cup \{h\})$.
     We infer that $F\cup \{h\}$ is blocking since $F\cup \{h\}$ is the unique most preferred coalition of $h$, a contradiction.} 

  Next, %
  we claim the following: %
  \newH{
    \begin{claim}\label{clm:zz-G}
      For each $z \in [\enn]$, if $A[z]\notin \Pi$, then $\hat{A}[z] \cup \{\gamma[z-1]\}\in \Pi$.
    \end{claim}
    
    \begin{proof}[Proof of \cref{clm:zz-G}]
      \renewcommand{\qedsymbol}{$\diamond$}
      The proof is similar to the one for \cref{clm:zz-1}.
      Consider an arbitrary index~$z \in [\enn]$ and assume that $A[z]\notin \Pi$.
      Notice that, by construction, coalition~$A[z]$ is the unique most preferred coalition of $\gamma[z]$; recall that $\gamma[\enn]=w_j^i$.
      Then, to prevent $A[z]$ from blocking~$\Pi$,
      at least one agent from~$\hat{A}[z]$ has to be matched in a coalition that she weakly prefers to~$A[z]$.
      Again, by construction, since coalition~$\hat{A}[z]\cup \{\gamma[z-1]\}$
      is the only coalition which is weakly preferred to~$A[z]$ by $a$ for some~$a\in \hat{A}[z]$.
      Therefore, $\hat{A}[z] \cup \{\gamma[z-1]\} \in \Pi$. 
    \end{proof}
  }
  
  Using a reasoning similar to the one for \cref{lem:set-agent}, by
  the first statement %
  and \cref{clm:zz-G}, 
  we can obtain the second statement in the lemma.
  We prove here for the sake of completeness.
  { \begin{description}
    \item[\textbf{Case 1:} {$\hat{A}[\enn]\cup \{\gamma[\enn-1]\}\in \Pi$}.]
    This means that $A[\enn-1]\notin \Pi$.
    By \cref{clm:zz-G}, it follows that $\hat{A}[\enn-1] \cup \{\gamma[\enn-2]\}\in \Pi$.
    By repeatedly using the above reasoning, we infer that $\hat{A}[z]\cup \{\gamma[z-1]\}\in \Pi$ for all $z\in [\enn]$, as desired.
    \item[\textbf{Case 2:} {$\hat{A}[\enn]\cup \{\gamma[\enn-1]\} \notin \Pi$}.]
    Notice that $\hat{A}[\enn]\cup \{\gamma[\enn-1]\}$ is the unique most preferred coalition of each agent in $\hat{A}[\enn]$.
    To prevent $\hat{A}[\enn]\cup \{\gamma[\enn-1]\}$ from blocking, $\gamma[\enn-1]$ has to weakly prefer $\Pi(\gamma[\enn-1])$ to $\hat{A}[\enn]\cup \{\gamma[\enn-1]\}$.    
    By construction, $\Pi(\gamma[\enn-1])$ has to contain at least one agent from~$\hat{A}[\enn-1]$, implying that $\hat{A}[\enn-1]\cup \{\gamma[\enn-2]\} \notin \Pi$. 
    By the contra-positive of \cref{clm:zz-G},
    we infer that $A[\enn-1]\in \Pi$.
    By repeatedly using the above reasoning, we infer that $A[z]\in \Pi$ for all $z\in [\enn-1]$.
    It remains to show that $A[\enn]\in \Pi$.
    Since $A[\enn-1]\in \Pi$ meaning that $\gamma[\enn-1]$ is not available,
    by construction,
    we infer that $A[\enn]$ is the unique more preferred coalition of each agent in~$A[\enn]$,
    and hence by stability that $A[\enn]\in \Pi$.
  \end{description}}
This completes the proof of the second statement.
\end{proof}
}
Finally, we show that the subcollection~$\mathcal{K}$ with $\mathcal{K}=\{S_j \in \mathcal{S}\mid
\hat{A}^j_i[1] \cup \{\gamma^j_i[0]\} \in \Pi \text{ for some } i \in S_j\}$ is an exact cover.
First, we claim that $\mathcal{K}$ does not cover an element more than once. For each two chosen~$S_j, S_k\in \mathcal{K}$ we observe that it cannot happen that
$S_j\cap S_k\neq \emptyset$ as otherwise $\{u_i^j,u_i^k\} \cup U_i$ is a blocking coalition; recall that $\gamma^j_i[0] = u_i^j$ and $\gamma^k_i[0]= u_i^k$.

Second, we claim that $\mathcal{K}$ is a cover. For each element~$i\in X$, let $S_j,S_k,S_{\setind}$ denote the three sets that contain~$i$.
We claim that at least one of $S_j,S_k,S_{\setind}$ belongs to~$\mathcal{K}$ because of the following.
If $S_j\notin \mathcal{K}$, then by construction, it follows that $T=\hat{A}^j_i[1] \cup \{\gamma^j_i[0]\}\notin \Pi$.
By \cref{lem:set-agentG}\eqref{lem:left-rightG}, it follows that $A^j_i[1]\in \Pi$.
Since by construction $T$ is the only most-preferred  coalition for each agent of $\hat{A}^j_i[1]$,
by stability, $u_i^j$ must be matched in a coalition which she weakly prefers to~$T$; recall that $u^j_i=\gamma^j_i[0]$. 
Since $A^j_i[1]\in \Pi$, this means that $\{u^j_i,v\} \cup U_i \in \Pi$ for some agent $v$.
It cannot happen that  $ v\notin \{u_i^k,u^{\setind}_i\}$ as otherwise by construction there will be at least three blocking coalitions, including $U_i \cup \{u_i^j,u^{\setind}_i\}$.
Hence, $\{u^j_i,v\} \cup U_i \in \Pi$ for some $v\in \{u_i^k,u^{\setind}_i\}$.
Without loss of generality, assume that $v=u^k_i$.
Then, it is straightforward to check that $\{u^{\setind}_i \} \cup \hat{A}^{\setind}_i[1] \in \Pi$ since $\{u^{\setind}_i \} \cup \hat{A}^{\setind}_i[1] \in \Pi$ is the unique most preferred coalition of each agent in $\hat{A}^{\setind}_i[1]$.
This implies that $S_{\setind}\in \mathcal{K}$.

{Finally, we show that $\hat{A}^{\setind}_p[1]\cup \{\gamma^{\setind}_{p}[0]\} \in \Pi$ for all $p\in S_{\setind}$.
  Suppose, for the sake of contradiction, that $\hat{A}^{\setind}_p[1]\cup \{\gamma^{\setind}_{p}[0]\} \notin \Pi$ for some $p \in S_{\setind}$.
  By \cref{lem:set-agentG}\eqref{lem:left-rightG}, we infer that $A^{\setind}_p[\enn] \in \Pi$, and that $w_{\setind}^p=\gamma_{p}^{\setind}[\enn]$ is not available for $w_{{\setind}}^p$ (for some constant~$\enn$).
  Thus, $\{w_{\setind}^p\}\cup F_{{\setind}}^p$ forms a blocking coalition:
  By \cref{lem:set-agentG}\eqref{lem:fghG},  each agent in~$F_{{\setind}}^p$ prefers $F_{{\setind}}^p\cup \{w_{{\setind}}^p\}$ to its assigned coalition~$F_{{\setind}}^p\cup \{h_{{\setind}}^p\}$.
  Agent~$w_{\setind}^p$ also prefers $F_{{\setind}}^p\cup \{w_{{\setind}}^p\}$ to its assigned coalition since
  $\distP{w_{\setind}^p} \ge 17.5 + (\di-3)\cdot 10 + 20 > = (10+\frac{15}{\di-1})\cdot (\di-1) = \dist(w_{\setind}^p, F_{\setind}^p)$ (besides $W_{\setind}\cup \{w_{\setind}^q\}$ the next available agent is of distance at least $20$ to $w_{\setind}^p$).
}
Together we infer that $\mathcal{K}$ is indeed an exact cover.
This concludes the proof for all $\di \ge 4$ and the proof for \cref{thm:main}.
}
\section{Conclusion and Outlook}\label{sec:conclude}

Establishing the first complexity results in the study of multi-dimensional stable matchings for Euclidean preferences, we show that \dsr{} remains NP-hard for Euclidean preferences and for all fixed~$\di\ge 3$.
The gadgets in the reductions may be useful for other matching and hedonic games problems with Euclidean preferences.

Our Euclidean preference model assumes that the preferences over coalitions are based on the sum of distances to all individual agents in the coalition.
It would be interesting to see whether taking the maximum or the minimum distance to the coalition members instead of the sum would change the complexity.
Furthermore, it would be interesting to see whether restricting the agents' embedding to $1$-dimensional Euclidean space could lower the complexity.
We were not able to identify the complexity for this restricted variant,
but conjecture that it can be solved in polynomial time.
Note that in $1$-dimensional Euclidean space,
 a \dsm[3] for the maximum distance setting always exists, which can be found by greedily finding three consecutive agents which are closest to each other and matching them. 

\clearpage

\bibliographystyle{plainurl}%

\bibliography{bib}

\clearpage
\appendix
\section*{Appendix}
\appendixProofText

\end{document}